\documentclass[11pt,a4paper]{article}
\pdfoutput=1
\usepackage{jheppub}
\usepackage{slashed}
\usepackage{multirow}
\usepackage{color}
\usepackage{mathtools}

\newcommand{\slv}{\raise.15ex\hbox{$/$}\kern-.53em\hbox{$v$}}
\newcommand{\slF}{\raise.15ex\hbox{$/$}\kern-.53em\hbox{$F$}}
\newcommand{\slL}{\raise.15ex\hbox{$/$}\kern-.53em\hbox{$L$}}
\newcommand{\slP}{\raise.15ex\hbox{$/$}\kern-.53em\hbox{$P$}}
\newcommand{\slp}{\raise.15ex\hbox{$/$}\kern-.53em\hbox{$p$}}
\newcommand{\slq}{\raise.15ex\hbox{$/$}\kern-.53em\hbox{$q$}}
\newcommand{\slR}{\raise.15ex\hbox{$/$}\kern-.53em\hbox{$R$}}
\newcommand{\slQ}{\raise.15ex\hbox{$/$}\kern-.53em\hbox{$Q$}}
\newcommand{\slK}{\raise.15ex\hbox{$/$}\kern-.53em\hbox{$K$}}
\newcommand{\slk}{\raise.15ex\hbox{$/$}\kern-.53em\hbox{$k$}}
\newcommand{\slD}{\raise.15ex\hbox{$/$}\kern-.53em\hbox{$D$}}
\newcommand{\slC}{\raise.15ex\hbox{$/$}\kern-.53em\hbox{$C$}}
\newcommand{\slA}{\raise.15ex\hbox{$/$}\kern-.53em\hbox{$A$}}
\newcommand{\slSigma}{\raise.15ex\hbox{$/$}\kern-.53em\hbox{$\Sigma$}}
\newcommand{\slpartial}{\raise.15ex\hbox{$/$}\kern-.53em\hbox{$\partial$}}
\newcommand{\slcalP}{\raise.15ex\hbox{$/$}\kern-.63em\hbox{$\cal P$}}

\def\bs{\boldsymbol}
\def\del{\partial}
\def\bdel{\boldsymbol \partial}

\def\p{{\boldsymbol p}}

\def\pp{{\boldsymbol p}}
\def\q{{\boldsymbol q}}
\def\l{{\boldsymbol l}}
\def\k{{\boldsymbol k}}

\def\x{{\boldsymbol x}}
\def\y{{\boldsymbol y}}

\def\r{{\boldsymbol r}}
\def\z{{\boldsymbol z}}

\def\u{{\boldsymbol u}}

\def\b{{\boldsymbol b}}

\def\G{{\cal G}}
\def\U{\mathcal{U}}
\def\A{\mathcal{A}^-_{med}}
\def\J{\mathcal{J}}

\def\P{\mathcal{P}}
\def\bkappa{{\boldsymbol \kappa}}
\def\bbkappa{\bar{\boldsymbol \kappa}}

\def\KO{{\cal K}_{osc}}

\def\S{{\cal S}}
\def\SO{{\cal S}_{osc}}
\def\bpar{{\boldsymbol \partial}}

\def\J{\mathcal{J}}

\newcommand{\beq}{\begin{eqnarray}}
\newcommand{\eeq}{\end{eqnarray}}
\newcommand{\be}{\begin{eqnarray*}}
\newcommand{\ee}{\end{eqnarray*}}

\newcommand{\nn}{\nonumber\\ }

\title{Coherence phenomena between initial and final state radiation in a dense QCD medium}
\author[a,b]{N\'estor Armesto,}
\author[a]{Hao Ma,}
\author[a,b]{Mauricio Mart\'inez,}
\author[c]{Yacine Mehtar-Tani,}
\author[a]{and Carlos A. Salgado}
\affiliation[a]{Departamento de F\'isica de Part\'iculas and IGFAE, 
Universidade de Santiago de Compostela, 
E-15706 Santiago de Compostela, 
Galicia-Spain}
\affiliation[b]{Physics Department, Theory Unit, CERN, CH-1211 Gen\`eve 23, Switzerland}
\affiliation[c]{Institut de Physique Th\'eorique, CEA Saclay, 
F-91191 Gif-sur-Yvette, France}

\preprint{CERN-PH-TH/2013-171}

\emailAdd{nestor.armesto@usc.es}
\emailAdd{hao.ma@usc.es}
\emailAdd{mauricio.martinez@usc.es}
\emailAdd{yacine.mehtar-tani@cea.fr}
\emailAdd{carlos.salgado@usc.es}

\abstract{
In this work we investigate medium modifications to  the interference pattern between initial and final state radiation.  We compute single gluon production off a highly energetic parton that undergoes a hard scattering and subsequently crosses a dense QCD medium of finite size. We extend our previous studies obtained at first order in opacity by providing general results for multiple soft scatterings and their specific formulation within the harmonic oscillator approximation.  We show that there is a gradual onset of decoherence between the initial and final state radiation due to multiple scatterings, that opens the phase space for large angle emissions. By examining the multiplicity of produced gluons, we observe a potentially large double logarithmic enhancement for dense media and small opening angles. This result points to a possible modification of the evolution equations due to a QCD medium of finite size. We briefly comment on the phenomenological consequences of this setup in high-energy nuclear collisions.  
}

\begin{document}
\maketitle

\section{Introduction}
\label{sec:intro}
Highly energetic jets emerging from the produced QCD matter are used as a diagnosis tool to extract information about the stages of ultrarelativistic heavy-ion collisions. Modifications of the space-time evolution of the jet compared to that in vacuum, are expected. Such modifications have already been experimentally confirmed by measurements of single inclusive particle spectra \cite{Adler:2003au,Adams:2003kv,Adams:2003im,Aamodt:2010jd,CMS:2012aa} and two-particle correlations \cite{Adler:2005ee,Adams:2005ph,Aamodt:2011vg} at RHIC and the LHC. Another manifestation of the same physics has been observed in jet studies in PbPb collisions at the LHC \cite{Aad:2010bu,Aad:2012vca,Chatrchyan:2011sx,Chatrchyan:2012nia,Chatrchyan:2012gt,Chatrchyan:2012gw}  (see also \cite{Salur:2008hs,Lai:2009zq} for related results at RHIC).  Summarizing, the results from jet measurements at the LHC show: (i) a larger imbalance of the transverse energy of leading and subleading jets in PbPb collisions than in $pp$ and increasing with centrality, indicating the presence of medium-induced energy loss; (ii) an  azimuthal distribution between leading and subleading jets in central PbPb collisions that turns out to be similar to that in $pp$, suggestive of the absence of sizeable medium-induced broadening in transverse momentum; (iii) an excess, compared to Monte Carlo expectations which reproduce $pp$ data, of soft particles at large angles with respect to the subleading jet in PbPb collisions and increasing with increasing dijet momentum imbalance; (iv) a lack of sizeable modifications, between $pp$ and PbPb collisions, of the jet fragmentation into particles with energies close to the jet energy. At first sight, these observations challenge the standard explanation of jet quenching in terms of medium-induced gluon radiation \cite{CasalderreySolana:2007zz,Wiedemann:2009sh,Majumder:2010qh} in which energy loss and broadening are linked and the induced radiation is semihard. Further studies are thus required to understand all these aspects.  

An important ingredient in a jet vacuum parton shower -- the Markovian process of subsequent parton splittings implemented e.g. in Monte Carlo codes to describe the jet evolution -- is the role of colour coherence. The basic question addressed by colour coherence is to what extent different emitters in the cascade act independently and the simplest setup to address this question is the antenna radiation. Considering the case of a quark-antiquark pair in colour singlet configuration with an opening angle $\theta_{q\bar q}$, colour coherence will suppress the radiation for soft gluons at emission angles $\theta>\theta_{q\bar q}$ while the radiation is basically unmodified at $\theta<\theta_{q\bar q}$. This simple example is the basis of the {\it angular ordering} prescription in jet evolution \cite{Dokshitzer:1991wu,Ellis:1991qj,Bassetto:1984ik}. When the antenna is a non-singlet colour state, radiation outside the cone, $\theta>\theta_{q\bar q}$, happens with the strength given by the total colour charge i.e. the pair acts as a single emitter with the charge of the pair. In this sense, a simple probabilistic picture emerges, which is implemented e.g. in Monte Carlo generators: radiation inside the cone  determined by the pair opening angle, $\theta<\theta_{\rm q\bar q}$,  takes place as independent radiation off each of the partons, while radiation out of the cone, $\theta>\theta_{\rm q\bar q}$, can be reinterpreted as radiation off the parent parton, so that the angular ordering can be effectively introduced. Perhaps the most clear manifestation of this  phenomenon is the suppression of soft particles in the jet fragmentation function (the hump-backed plateau) that has been observed experimentally \cite{Braunschweig:1990yd,Abbiendi:2002mj,Abe:1994nj}. 
These interferences, and the subsequent angularly ordered pattern, occur not only between final state emitted partons in  time-like parton showers but also in space-like ones and between initial and final state emissions as the ones considered here.

The study of the fate of colour coherence when the branching process occurs inside a coloured medium has started recently \cite{MehtarTani:2010ma,CasalderreySolana:2011rz,MehtarTani:2011gf,MehtarTani:2012cy,Blaizot:2012fh,Fickinger:2013xwa}, commonly using an in-medium colour antenna as laboratory \cite{MehtarTani:2010ma,CasalderreySolana:2011rz,MehtarTani:2011gf,MehtarTani:2012cy}. The main findings of these studies are easy to understand in terms of a modification of the colour coherence of the two emitters in the antenna due to colour rotation in the medium. When the transverse size of the antenna  is smaller than the typical colour correlation length inside the medium, this cannot modify the internal colour structure of the antenna and  colour coherence is maintained: vacuum radiation is angularly ordered in the way described in the previous paragraph while the medium-induced radiation proceeds as if the antenna would be a single parton with the colour  of the pair. In the opposite limit, when the size of the antenna is similar or larger than the colour correlation length, the medium can  effectively decorrelate the pair by colour rotation so that each parton losses memory that it was once correlated with the other parton. In this situation, each parton emits medium-induced radiation independently and the angular ordering of the vacuum radiation is broken i.e. the two partons emit as two completely independent partons, both for medium-induced and for vacuum emissions. In the strict soft limit, when medium-induced radiation is suppressed, the name {\it antiangular ordering} was originally given to this effect in Ref. \cite{MehtarTani:2010ma}. The consequences for jet quenching in heavy-ion collisions are starting to be explored \cite{CasalderreySolana:2012ef}.

It looks natural to extend such studies to another situation different from final state emissions from an antenna. In our case, we focus on the interference between initial and final state radiation. We compute single gluon production off a highly energetic parton that undergoes a hard scattering and subsequently crosses a QCD medium of finite size. This setup was previously considered by us \cite{Armesto:2012qa} for the case of a finite size dilute medium, in which the interaction with it is modelled through a single hard scattering. In the soft limit, the two main conclusions about the full (medium + vacuum) gluon spectrum are: (i) a reduction of gluon emissions from the initial state by a quantity which depends on the medium properties but with the radiation confined inside the cone defined by the opening angle described by the initial and final emitting partons, and
(ii) final state emissions losing totally their vacuum coherence characteristics once in the medium, giving rise to large angle emissions
 (named antiangular ordering in the antenna studies) that arise from the medium-induced coherent radiation between both emitters. Both features resemble those found in the antenna, but they are based in an extension of the formalism to denser system that can only be properly addressed by considering the medium as a collection of many soft scatterers. It is to this extension that this paper is devoted. As a bonus, we gain insight into the different scales that rule gluon radiation from such a system, and a connection with other existing formalisms.
 
The paper is organised as follows: In Section \ref{sec:vac-coh} we briefly discuss the role of colour coherence in deep inelastic scattering (DIS)  which corresponds to the setup that we address in this work. In Section \ref{sec:amplitude} we revise the semiclassical formalism for gluon radiation \cite{Gelis:2005pt,Blaizot:2004wu,Iancu:2003xm} and calculate the relevant amplitudes both in vacuum and in medium. The reader familiar with these methods can skip this Section and go directly to Section \ref{sec:tspectrum} where the medium averages are computed to obtain the spectrum, which is then approximated for a harmonic oscillator. In Section \ref{sec:ftan}, a formation time analysis is performed to obtain the relevant scales and two possible limits, coherent and incoherent, are discussed.
In Section \ref{sec:discussion} we comment on possible modifications to the usual perturbative evolution by analysing the leading logarithmic behaviour of the gluon multiplicity. Finally we conclude with a summary and some remarks on possible phenomenological consequences of this setup in high-energy nuclear collisions.

\section{A few remarks on  colour coherence}
\label{sec:vac-coh}

Coherence phenomena in particle production is an important subject of study in high-energy particle collisions. Their effect has been confirmed experimentally in $e^+e^-$ annihilation \cite{Braunschweig:1990yd,Abbiendi:2002mj} and $p\bar{p}$ collisions \cite{Abe:1994nj}. This success has established the role of the non-abelian nature of QCD and the general validity of the perturbative approach for jet physics. The process of the cascading of a jet initiated by a highly energetic parton is Markovian i.e. multiple branchings in the shower generated by the parent parton are not independent: the angle of emission of a subsequent soft gluon is smaller than the one produced previously. Therefore, multiple soft gluons in a parton shower are emitted in a coherent manner since they are angularly ordered.   

To illustrate in a simple manner the role of angular ordering in a parton shower, let us consider soft gluon production in a DIS process of a highly energetic parton where there is no colour transfer in the $t$-channel exchange (e.g. an electromagnetic quark scattering)  at finite angle $\theta_{qq}$\footnote{In what follows we denote the $t$-channel exchange as the \textit{hard scattering}.} between the incoming and outgoing quark. For completeness, we perform the calculation of the gluon spectrum for this example in Appendix \ref{sec:vac-spec}. Gluon emission takes place either before or after the hard scattering off the incoming parton (when it is space-like) or  the outgoing parton (when it is time-like) respectively. Thus one expects to observe two cones of radiation centered along the longitudinal directions of any of the emitters. Every cone of radiation has an opening angle $\Theta\equiv \theta_{qq}$. This can be easily argued in the following way: If the transverse wavelength of the radiated gluon $\lambda_\perp\sim 1/k_\perp$ is smaller than the transverse displacement $r_\perp\sim \tau_{form}\theta_{qq}$ between the emitters at the time the gluon is emitted (the gluon formation time $\tau_{form}\sim \omega/k_\perp^2$ with $\omega$ and $k_\perp$ the energy and transverse momentum of the gluon respectively), the gluon is able to resolve the transverse distance between both emitters. If $\lambda_\perp > r_\perp$, the gluon does not measure effectively any transverse displacement so it would look as if the hard scattering did not affect at all the trajectory of the emitting parton. Thus, soft gluon emissions are absent at  large angles $\theta_k > \theta_{qq}$.  In the case of colour transfer in the $t$-channel there will be an additional contribution to the gluon spectrum at large angles $\theta_k > \theta_{qq}$, which indicates that even in the case of a negligible transverse displacement the incoming parton is affected by the $t$-channel colour exchange. This argument holds for multiple soft gluon emissions at any order in the coupling constant and in any colour configuration \cite{Dokshitzer:1991wu,Bassetto:1984ik,Ellis:1991qj}.

\section{Scattering amplitude from Classical Yang-Mills Equations}
\label{sec:amplitude}

In this section we outline the calculation of the scattering amplitude using  semiclassical methods in pQCD. We also introduce the notation followed through this work. This theoretical framework has shown to be an efficient tool to calculate inclusive observables involving soft gluon emissions at high energies \cite{Gelis:2005pt,Blaizot:2004wu,Iancu:2003xm}. Within this approach, a soft gluon is a solution of the linearized Classical Yang-Mills (CYM) equations of motion in the presence of a background field (the target) and a colour source which is a parton with large momentum (the projectile).  Through this work, we limit ourselves to the eikonal approximation which is valid as far as the gluon is soft relative to the parent partons i.e. $k^+\ll p^+,\bar{p}^+ $\footnote{Any 4-vector $x\equiv(x^0,x^1,x^2,x^3)$ in Minkowski space is described in  light-cone (LC) coordinates as $x\equiv(x^+,x^-,\x)$, where $x^\pm\equiv(x^0 \pm x^3)/\sqrt{2}$ and $\x = (x^1,x^2)$. For instance, in the LC coordinates the momentum of a particle $p^\mu=(p^+,p^-,{\bf p})$ where $p^{\pm}=(E\pm p_z)/\sqrt{2}$ being $E$ and $p_z$ the energy and the longitudinal momentum respectively.}.  In addition, we consider the region of small angles defined by  $p^+, \,\bar p^+\gg |\p|,\,|\bar \p|\gg k^+\gg |\k|$ which is the interesting region in intrajet physics, see Fig. \ref{fig:contr-ampl} for the notation.

Let us recall first the standard reduction formula which relates the amplitude for emitting a gluon with momentum $k^\mu\equiv(k^+,k^-=\k^2/(2k^+),\k)$ with the classical gauge field $A_\mu^a$ (see for instance \cite{Itzykson:1980rh}):
%
\beq
\label{eq:redform-a}
{\cal M}_\lambda ^{a}({k})&=& \lim_{k^2\to 0 } -k^2\,A^{a}_\mu(k)\epsilon^\mu_\lambda({ k})\,,
\nonumber \\
&=&\lim_{k^2\to 0 } \int d^4x\, e^{ik\cdot x}\, \square_x A^{a}_\mu(x)  \epsilon^\mu_\lambda({ k}) \,,
\eeq
where $\epsilon^\mu_\lambda({k})$ is the gluon polarization vector. The classical gauge field $A^a_\mu$ is the solution of the CYM equations
\beq
\label{eq:CYM}
[D_\mu,F^{\mu\nu}] = \mathcal{J}^\nu, 
\eeq
with $D_\mu\equiv \del_\mu-ig A_\mu$ and $F_{\mu\nu}\equiv \del_\mu A_\nu-\del_\nu A_\mu-ig[A_\mu,A_\nu]$. In addition to the CYM Eqs. (\ref{eq:CYM}), one must consider the continuity equation for the classical colour current $[D_\mu,\mathcal{J}^\mu]=0$ which describes the space-time evolution of the projectile. We concentrate on asymptotic states far from the region where the physical process happens i.e. at $x^+\to\infty$. Therefore, Eq. (\ref{eq:redform-a}) can be rewritten as \cite{MehtarTani:2011jw,Blaizot:2004wu}
\beq
\label{eq:redform2}
{\cal M}_\lambda ^{a}({ k})= \lim_{x^+\to \infty } 
\int dx^-d^2\x\, e^{ik\cdot x}\, 2
\frac{\partial}{\partial x_+} 
{\bs A}^{a}(x) \cdot {\bs \epsilon}_\lambda(\vec k)  \,.
\eeq
We perform our calculations in the light-cone gauge (LCG) $A^+=0$. Then, only the transverse polarization contributes to the radiative cross-section since $\sum_{\lambda} \epsilon^i_\lambda(\epsilon_\lambda^j)^\ast=\delta^{ij}$, where $i(j)=1,2$. In the LCG the gluon polarization vector $\epsilon^\mu_\lambda=(0,{\boldsymbol \epsilon}_\lambda\cdot\k/k^+,{\boldsymbol \epsilon}_\lambda)$. Finally, the gluon spectrum reads
\beq
\label{eq:gluspec}
(2\pi)^3 2k^+ \frac{dN}{d^3k} =\sum_{\lambda=1,2} |{\cal M}_\lambda ^{a}({\vec k})|^2 \,,
\eeq
where the phase space volume in momentum space is $d^3k\equiv d^2\k \,dk^+$.

\subsection{Gluon emission amplitude for the initial and final state radiation}
\label{subsec:ampl}

%
\begin{figure}[t]
\begin{center}
\includegraphics[width=14cm]{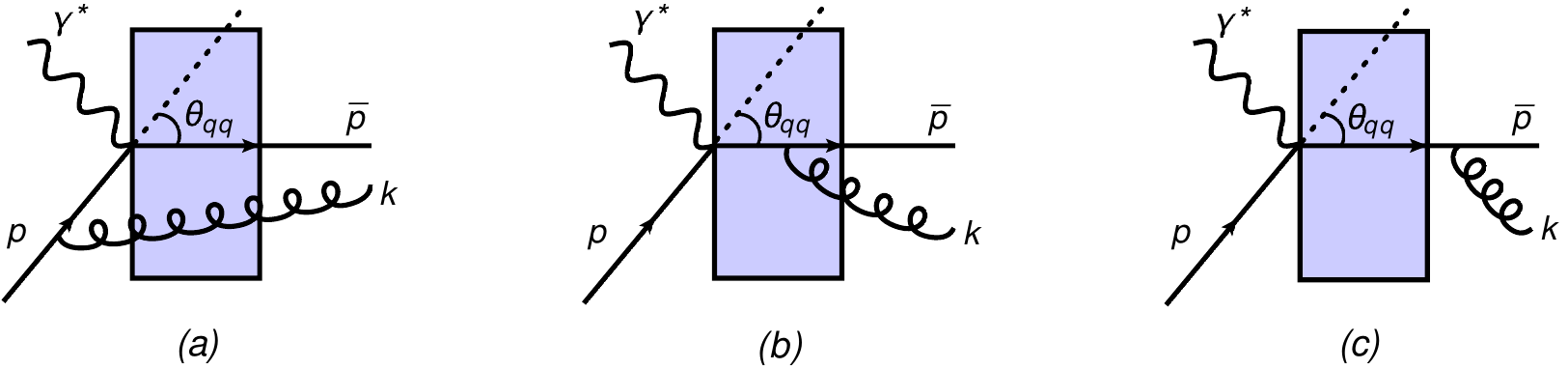}
\end{center}
\caption{Different contributions to the scattering amplitude. The gluon can be radiated either before (a) or after the hard scattering ((b) and (c)). In the later case, the emission can take place either inside (b) or outside the QCD medium (c).}
\label{fig:contr-ampl}
\end{figure}

By making use of the semiclassical method described previously, we calculate the single inclusive gluon spectrum off a highly energetic parton created in the remote past which suffers a hard scattering at $x^+_0=0$ and subsequently passes through a dense QCD medium of finite longitudinal size $L^+$.
Gluon emission takes place either before or after the hard scattering (see Fig. \ref{fig:contr-ampl}). The QCD medium is modelled by a background gauge field $\mathcal{A}^{a,-}_{med}$ which is a solution of the two dimensional Poisson equation $-\partial_\x A^{a,-}_{med}=\rho^a(x^+,\x)$, where $\rho^a(x^+,\x)$ is the static distribution  of medium colour charges. 

In the vacuum case the total classical eikonalized current that describes the projectile either before ($bef$) or after ($aft$) the hard scattering is $\J^{\mu}_{(0)}=\J^\mu_{bef,(0)}+\J^{\mu}_{aft,(0)}$, where $\J^{\mu}_{bef,(0)}$ and $\J^{\mu}_{aft,(0)}$ are 
\begin{subequations}
\label{eq:inc-outcurr}
\beq
\label{eq:inc-curr}
\J^{\mu,a}_{bef,(0)}(x)&= g u^\mu \Theta (x^+_0-x^+)\,\delta (x^- - u^- x^+)\delta^{(2)}(\x-\u x^+) \, Q^a_{bef}\,, \\
\label{eq:out-curr}
\J^{\mu,a}_{aft,(0)}(x)&= g \bar{u}^\mu \Theta (x^+-x^+_0)\,\delta (x^- - \bar{u}^- x^+)\delta^{(2)}(\x-\bar{\u} x^+)\, Q^a_{aft}\,.
\eeq
\end{subequations}
In the last expression we have used the definition of the 4-velocity for each parton in LC coordinates $u^\mu= p^\mu/p^+\equiv (1,u^-,\u)$ and $Q^a_{bef(aft)}$ denotes the colour charge of the incoming (outgoing) parton\footnote{ An overlining $\bar{\ \ }$ is used hereafter to denote  quantities related to the outgoing parton.}.  In the vacuum case the colour current conservation implies $Q^a_{bef}=Q^a_{aft}$ and the scalar product of two colour charges is given by $ Q_i Q_j = C_F$ for the colour singlet case. Note that the gluon spectrum for the vacuum can be obtained using Eqs. (\ref{eq:inc-outcurr}) (see Appendix \ref{sec:vac-spec}). 

To calculate the net effect over the total colour current $\J^\mu$ when there is a QCD medium one must solve the continuity equation $[D_\mu,\mathcal{J}^\mu]=0$ when $\A(x^+,\x)\ne 0$. The solution for the total colour current is, as in the vacuum case, $\J^\mu=\J^{\mu}_{bef}+\J^{\mu}_{aft}$ where
\begin{subequations}
\label{eq:med-curr}
\beq
\label{eq:med-inccurr}
\J^{\mu,a}_{bef}(x)&=& \J^{\mu,a}_{bef,(0)}(x)\,,\\
\label{eq:med-outcurr}
\J^{\mu,a}_{aft}(x)&=&\U^{ab}(x^+,0) \J^{b,\mu}_{aft,(0)}(x)\,.
\eeq
\end{subequations}
In Eq.(\ref{eq:med-outcurr}) $\U^{ab}(x^+,0)$ is a Wilson line in the adjoint representation whose general definition is
\beq
\label{eq:WL}
\U^{ab}(x^+,y^+)={\cal P}\exp\left[ig\int _{y^+}^{x^+} \!\!dz^+\, \A\left(z^+,\r(z^+) \right)\right] ^{ab}\,,
\eeq
where ${\cal P}$ denotes path ordering, $\r(z^+)$ is the trajectory of the probe along the transverse path (e.g. it reads $\r=\u z^+$ for a parton moving with constant velocity $\u$ in the transverse plane), and $\A=T^a {\cal A}^{a,-}_{med}$.  The solution of the continuity equation for the total colour current has a simple interpretation: before the hard scattering the classical current $\J^{\mu}_{bef}$ does not interact with any of the constituents of the QCD medium while the eikonalized current $J^\mu_{aft}$  gets colour-rotated due to multiple scatterings with the background field $\mathcal{A}^-_{med}$. The colour rotation between the medium and the colour current $J^\mu_{aft}$ is taking into account through the Wilson line (\ref{eq:WL}). Conservation of colour charge is still satisfied ($Q^{bef}=Q^{aft}$) since the presence of the medium does not change the values of the total colour charge of the projectile while passing through the medium, as expected \cite{Gelis:2005pt,Blaizot:2004wu,MehtarTani:2006xq}. 

Now that we know the solution of the total colour current $\J^\mu$ in the presence of $\A$, we must solve and linearize the CYM Eqs. (\ref{eq:CYM}) for the total gauge field $A^{\mu}=\delta^{\mu -}\A+a^\mu$, with $a^\mu$ a small perturbation around the background field $\A$. The linearized version of the CYM Eqs. (\ref{eq:CYM}) in the LCG\footnote{Hereafter, contraction of colour indices is to be understood when appropriate.} is given by \cite{Gelis:2005pt,MehtarTani:2006xq}
\begin{subequations}
\beq
\label{eq:CYM-constraint}
&\partial_- a^-+\partial_i a^i =-\frac{\J^+}{\partial_-}\,,\\
\label{eq:CYM-a}
&\square_x a^--2i g \bigl[\A , \partial_- a^-\bigr] - 2i g \bigl[a^i,\partial_i \A\bigr] + ig [\A, \partial_- a^-+\partial_i a^i]= \J^-\,,\\
\label{eq:CYM-ai}
&\square_x a^i-2i g \bigl[\A,\partial_- a^i\bigr] = \J^i-\partial^i \Bigl(\frac{\J^+}{\partial_-}\Bigr)\,.
\eeq
\end{subequations}
We concentrate only on the transverse components $a^i$ since these are the only ones that contribute to the scattering amplitude in the LCG\footnote{In principal one can also find the component $a^-$ either by solving directly  Eq. (\ref{eq:CYM-a}) or use the constraint Eq. (\ref{eq:CYM-constraint}) provided a known solution for $a^i$. See Refs. \cite{Gelis:2005pt,Blaizot:2004wu} where this procedure was done in the CGC context.}. The solution of Eq. (\ref{eq:CYM-ai}) is given by \cite{Blaizot:2004wu,MehtarTani:2006xq} 
\beq
\label{eq:aisol0}
a^i_a(x^+,\x,k^+)= \int d^4y\, G_{ab}(x,y)\tilde{\J}^i_b(y)
\eeq
where the modified current reads
\beq
\tilde{\J}^i=\J^i-\partial^i \Bigl(\frac{\J^+}{\partial_-}\Bigr)
\eeq
and $G_{ab}$ is the retarded Green's function of the differential equation
\beq
\square_x  G_{ab}(x,y)- 2\,ig\bigl[\A, \partial^+ G(x,y)\bigr]_{ab}=\delta_{ab}\delta^{(4)}(x-y).
\eeq
The background field does not depend on $x^-$, thus the Green's function $G_{ab}(x,y)$ is invariant under translations along this direction. It is convenient to introduce here a Fourier transform of the Green's function $G_{ab}$
\beq
{\cal G}_{ab}(x^+,\x\,;\,y^+,\y|k^+)=\int^{+\infty}_{-\infty}dx^- e^{i(x-y)^- k^+} 2
\frac{\partial}{\partial x_+} 
G_{ab}(x,y) \,,
\eeq
which obeys the Schr\"odinger-like equation
\beq
\left(i\del^-+\frac{\bdel^2}{2k^+}\right)\,\G_{ab}(x^+,\x\,;\,y^+,\y|k^+)&+&g\, \bigl[\A,\mathcal{G}(x^+,\x\,;\,y^+,\y|k^+)\Bigr]_{ab} \nonumber \\
&=&i\delta_{ab}\delta(x^+-y^+)\delta(\x-\y)
\eeq
($\bdel$ denotes the gradient in the transverse coordinates) and its solution is written as a path integral along the transverse plane \cite{MehtarTani:2006xq}
\beq
\mathcal{G}_{ab}\left(x^+,\x; y^+,\y| k^+ \right) = \int_{\r(y^+)=\y}^{\r(x^+)=\x}\mathcal{D} {\bs r} \, \exp\left[i\frac{k^+}{2}\int_{y^+}^{x^+} \!\!d\xi \, \dot{{\bs r}}^2(\xi) \right] \U_{ab}(x^+,y^+) \,,
\eeq
with $\dot{{\bs r}}(\xi)=\frac{d {{\bs r}}(\xi)}{d\xi}$.
This propagator takes into account the non-eikonal corrections to the emitted gluon due to the momentum broadening acquired due to the multiple scatterings with the medium. 
Finally, by taking the solution of the radiation field $a^i$ (\ref{eq:aisol0}) and replacing it into Eq. (\ref{eq:redform2}), we get the gluon emission amplitude
\beq
\label{eq:totamplitude0}
{\cal M}_{tot, \lambda} ^{a}({k})= \lim_{x^+\to \infty } \int d^2\x\,d^4y\, e^{i (k^-x^+-\k\cdot\x)} \, e^{ik^+y^-} \G_{ab}\left(x^+,\x; y^+,\y| k^+ \right)\tilde{\J}^b(y)\cdot {\boldsymbol \epsilon}_\lambda\,.
\eeq
By knowing the solution for the total colour current $J^\mu$ and using the linearity of the scattering amplitude with respect to the colour sources, we can split ${\cal M}_{tot}$ into two pieces that we interpret as the contribution of the incoming and the outgoing parton respectively. In the rest of this section we show the explicit form of every contribution. 

\subsubsection{The outgoing contribution to the scattering amplitude}
\label{subsec:outscatt}

Depending on the longitudinal position $y^+$ where the gluon is emitted, the scattering amplitude of the outgoing parton can be splitted in two pieces: when $y^+ \in$ $[0,L^+]$ the emission occurs inside the medium ($in$) and when $y^+\geq L^+$ the emission takes place outside ($out$) the medium (See Fig. \ref{fig:contr-ampl}). This is easily achieved by separating the integral over $y^+$ as 
\begin{equation*}
\int_0^\infty dy^+=\int_0^{L^+}dy^++\int_{L^+}^\infty dy^+.
\end{equation*} 
Then, one replaces the colour current (\ref{eq:med-outcurr}) into Eq. (\ref{eq:totamplitude0}) and after some algebra, the scattering amplitude associated to the outgoing quark current reads as ${\cal M}_{\lambda, aft} ^{a}({ k})= {\cal M}_{\lambda, in} ^{a}({ k})+{\cal M}_{\lambda, out} ^{a}({k})$ where \cite{CasalderreySolana:2011rz,MehtarTani:2006xq,MehtarTani:2011jw}
\begin{subequations}
\label{eq:outampl}
\beq
\label{eq:outampl-ins}
{\cal M}_{\lambda, in} ^{a}({k})&=&\frac{g}{k^+}\, \int d^2\x e^{i (k^-L^+-\k\cdot\x)}  \int_{0}^{L^+} dy^+ e^{i k^+ \bar{u}^-y^+}\\
&\times&\, {\boldsymbol \epsilon}_\lambda\cdot\bigl(i\bpar_y +k^+\bar{\u}\bigr)\mathcal{G}_{ab}\bigl(L^+,\x,y^+,\y=\bar{\u}y^+| k^+ \bigr)\,\U_{bc}(y^+,0)Q^{out}_c\ ,\nonumber\\
\label{eq:outampl-out}
{\cal M}_{\lambda, out} ^{a}({k})&=&-2i \frac{{\boldsymbol \epsilon}_\lambda\cdot\bbkappa}{\bbkappa^2}\, e^{i (k\cdot\bar{\u})L^+} \,\U_{ab}(L^+,0)Q^{out}_b\ ,
\eeq
\end{subequations}
where we introduce the transverse vector $\bar{\kappa}^i = k^i - \bar{x}\,\bar{p}^i$, $i=1,2$. This vector describes the transverse momentum of the gluon relative to the one of the outgoing quark. We define the momentum fraction carried out by the emitted gluon with respect to the outgoing parton $\bar{x}=\bar{k}^+/\pp^+$ (a similar definition follows for the incoming parton). Moreover, it must be understood in Eq. (\ref{eq:outampl-ins}) that, after the performing of the transverse derivatives $\partial_\y$, one sets $\y=\bar{\u}y^+$. 

The emission amplitude inside the medium ${\cal M}_{in}$, Eq. (\ref{eq:outampl-ins}), can be understood as a two-step process: initially the highly energetic parton gets colour precessed from $0$ until $y^+$ where a gluon is emitted; afterwards, the radiated gluon experiences transverse momentum broadening  until the edge of the medium at $L^+$. The physical information regarding the broadening of the radiated gluon is encoded in the retarded propagator $\mathcal{G}$. In the case of ${\cal M}_{out}$, Eq. (\ref{eq:outampl-out}), the projectile is rotated in colour along the total length of the medium and the gluon is emitted by bremsstrahlung outside of the medium.

\subsubsection{The incoming contribution to the scattering amplitude}
\label{subsec:inscatt}

The gluon radiation off the incoming parton happens completely before  the hard scattering i.e. when $y^+\in[-\infty,0]$. 
The emission amplitude in this case is found after  replacing Eq. (\ref{eq:med-inccurr}) into Eq. (\ref{eq:totamplitude0}) so the contribution associated to the incoming parton reads
\beq
\label{eq:inampl-1}
{\cal M}_{\lambda, bef}^{a}({k})&=&\frac{g}{k^+}  \int_{x^+=\infty} d^2\x\, e^{i (k^-x^+-\k\cdot\x)} \int_{-\infty}^0 dy^+ e^{i k^+ u^-y^+}\nonumber\\
&\times&{\boldsymbol \epsilon}_\lambda\cdot\bigl(i\bpar_y +k^+\u\bigr)\mathcal{G}_{ab}\bigl(x^+,\x,y^+,\y=\u y^+| k^+ \bigr)\,Q^{in}_b\ .
\eeq
As we pointed out in the previous section, there is no colour precession in the classical current describing the incoming parton due to the absence of a QCD medium before the hard scattering. So ${\cal M}_{bef}$ takes into account the emission by bremsstrahlung of the gluon and its subsequent classical broadening contained in the propagator $\mathcal{G}$. 

We conclude this section by recalling that the total scattering amplitude is the sum of the contributions   ${\cal M}_{tot}= {\cal M}_{aft}+{\cal M}_{ bef}$, where ${\cal M}_{aft}={\cal M}_{in}+{\cal M}_{out}$ and ${\cal M}_{bef}$ are given by Eqs. (\ref{eq:outampl}) and (\ref{eq:inampl-1}) respectively.

\subsection{Medium averages}
\label{subsec:medav}

A necessary ingredient to calculate observables in high energy nuclear collisions is  the distribution of colour charges in the target. Since this information is not known from first principles, it is usually assumed that the background field $\mathcal{A}^-_{med}$ is distributed along the medium as a Gaussian white noise i.e.
\beq
\label{eq:gaussian}
\langle \mathcal{A}^{a,-}_{med} (x^+,{\boldsymbol q}) \mathcal{A}^{*b,-}_{med} (x'^+,{\boldsymbol q}') \rangle = \delta^{ab} n(x^+) \delta (x^+-x'^+)\delta^{(2)}({\boldsymbol q}-{\boldsymbol q}'){\mathcal V}^2({\boldsymbol q}),
\eeq
where ${\mathcal V}({\boldsymbol q})$ is the medium interaction potential and $n(x^+)$ is the volume density of scattering centers. ${\mathcal V}({\boldsymbol q})$  is usually taken as a Debye-screened Coloumb potential \cite{Baier:1998yf,Baier:1996kr,Baier:1996sk,Wiedemann:2000ez,Wiedemann:2000za,Gyulassy:2000fs,Gyulassy:2000er}. 

In the Gaussian approximation (\ref{eq:gaussian}) and for the present calculation, we need to evaluate the correlators $\langle {\cal G}{\cal U}^\dagger\rangle$ and $\langle {\cal G}{\cal G}^\dagger\rangle$. The two point function $\langle {\cal G}{\cal U}^\dagger\rangle$ is related with the quak-gluon system in the medium and it is given by
\beq
\label{eq:qgdip}
\frac{1}{N_c^2-1}&&\,\langle{\text{Tr}}\,{\cal G}(x^+,\x;y^+,\y|k^+)\,{\cal U}^\dagger(x^+,y^+) \rangle = \nonumber\\
&&e^{ik^+ \u\cdot \left[\bar\x(x^+)-\bar\y(y^+) \right]}e^{i k^+u^-(x^+-y^+)} {\cal K}\left(x^+,\bar\x(x^+)\,;\,y^+, \bar\y(y^+) |k^+\right)\;,
 \eeq
 where $\bar\x(x^+)= \x -\u x^+$ and $\bar\y(y^+)= \y -\u y^+$. The multiple scattering of the gluon with the medium is taken into account through the path integral ${\cal K}$  \cite{Wiedemann:1999fq,Wiedemann:2000ez,Wiedemann:2000za,Zakharov:1996fv,Zakharov:1998sv},
 \beq
 \label{eq:GUcorr}
{\cal K}(x^+,\x;y^+,\y|k^+)= \int_{\r(y^+)=\y}^{\r(x^+)=\x}\mathcal{D} {\bs r} \, \exp\left[\int_{y^+}^{x^+} \!\!d\xi \left(i\frac{k^+}{2} \dot{{\bs r}}^2(\xi) - \frac{1}{2} n(\xi) \sigma\left({\bs r}(\xi)\right) \right)\right] \,,
\eeq
which describes the Brownian motion of the gluon the transverse plane from $\r(y^+)=\y$ to $\r(x^+)=\x$. The correlator $\langle {\cal G}{\cal G}^\dagger\rangle$ is related with the medium average involving the gluon line element from $y^+$ to $x^+$ and reads
\begin{multline}
\label{eq:GGcorr}
\int d^2\z \int d^2\z' \frac{e^{-i\k\cdot(\z-\z')}}{N_c^2-1}\langle {\text{Tr}}\, {\cal G}(x^+,\z;y^+,\x|k^+){\cal G}^\dag(x^+,\z';y^+,\y|k^+)\, \rangle=\\
e^{-i\k\cdot(\x-\y)}\S(x^+,y^+;\x-\y)\,,
\end{multline}
where $\S (x^+,y^+,\x-\y)$ is  the scattering $S$-matrix for an octet dipole of fixed transverse size $\x-\y$ that is defined as 
\beq
\label{eq:dip}
\S (x^+,y^+;\x-\y)=\exp\left[-\frac{1}{2}\int_{y^+}^{x^+} d\xi\, n(\xi)\,\sigma(\x-\y)\right] \,.
\eeq

Both medium averages (\ref{eq:GUcorr}) and (\ref{eq:GGcorr}) depend on the dipole cross section $\sigma (\r)$ whose general expression reads
\beq
\sigma (\r) =\int \frac{d^2\q}{(2\pi)^2}{\mathcal V}({\boldsymbol q})\bigl[1-\cos(\r\cdot\q)\bigr]\,.
\eeq

Following the usual procedure implemented in previous works,  we will make use of the `harmonic oscillator approximation' where the product $n(\xi)\sigma (\r)\approx \frac{1}{2}\hat{q}\r^2$ \cite{Zakharov:1996fv,Zakharov:1998sv}. This approximation is valid in the limit of multiple soft scatterings. Here $\hat{q}$ is the medium transport coefficient which probes the accumulated transverse momentum squared per unit mean free path. The harmonic oscillator approximation allows to calculate analytically the path integral (\ref{eq:GUcorr}) for the case of a static and homogeneous medium\footnote{Within the harmonic oscillator approximation it is also possible to evaluate the path integral (\ref{eq:GUcorr}) for the case of an expanding medium. Details of this procedure can be found in Refs. \cite{Baier:1998yf,Salgado:2003gb,CasalderreySolana:2007zz}.},
\beq
\label{eq:Kosc}
{\cal K}_{osc}(x^+,\x;y^+,\y|k^+)=\frac{A}{\pi i}\,\exp \bigl[ i\, AB (\x^2+\y^2)-2iA\,\x\cdot\y\bigr]\,,
\eeq
where
\beq
\label{eq:A-Bcoeff}
A=\frac{k^+\Omega}{2\sin (\Omega\Delta\tau)}\,, \ \ 
B=\cos\left[\Omega\Delta\tau\right],\ \ \Omega=\frac{1-i}{2}\sqrt{\frac{\hat q}{k^+}}\ ,
\eeq
with $\Delta\tau=x^+-y^+$. In addition, the $S$-matrix for a dipole (\ref{eq:dip}) in the harmonic oscillator approximation takes a simple form
 \beq
 \label{eq:Sosc}
 \S_{osc}(y^+,x^+;\r)= \exp\Biggl(-\frac{1}{4}\hat{q}\r^2(y^+-x^+)\Biggr)\ .
 \eeq

\section{The medium-induced gluon spectrum}
\label{sec:tspectrum}
%
\begin{figure*}[t]
\begin{center}
\includegraphics[width=15cm]{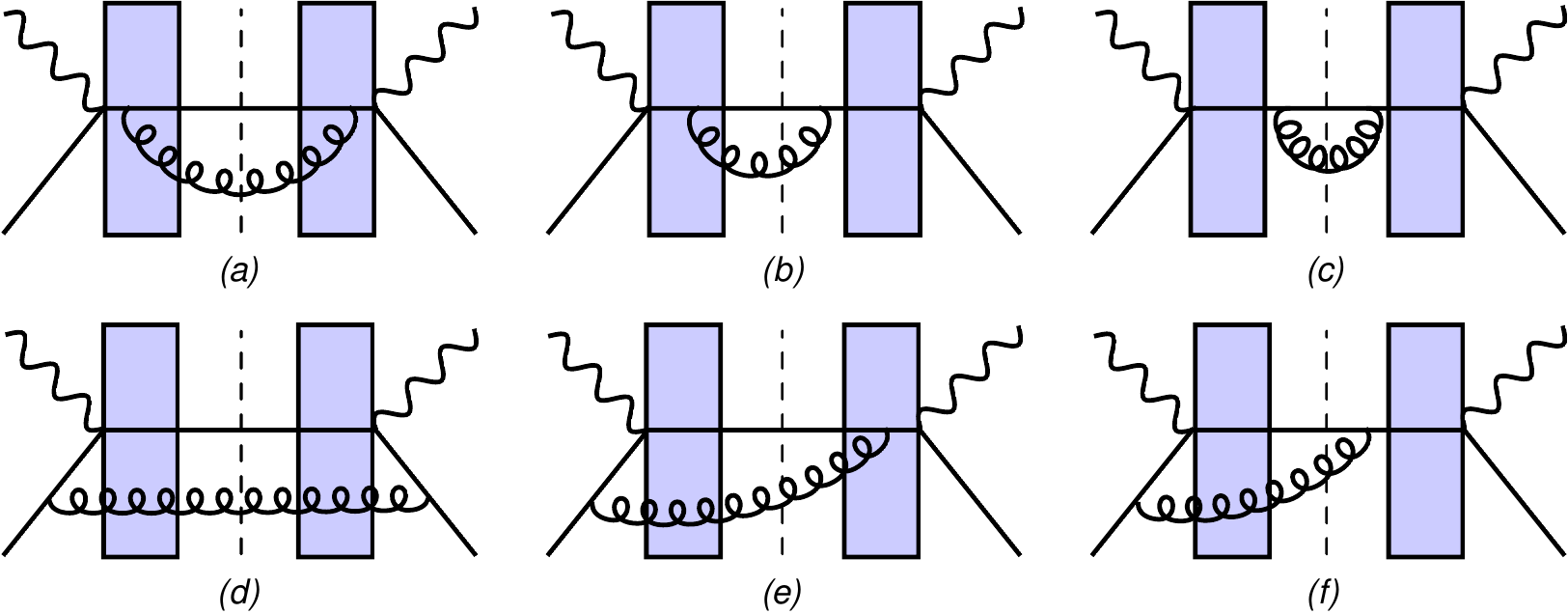}
\end{center}
\caption{Different components of the gluon spectrum when a QCD medium is present: (a) the ``\emph{in-in}" component, (b) the ``\emph{in-out}" component, (c) the ``\emph{out-out}" component, (d) the ``\emph{bef-bef}" component, (e) the ``\emph{bef-in}" component, and (f)  the ``\emph{bef-out}" component. The dashed line represents the cut which divides the amplitude on the left and its complex conjugate on the right. See the text for further details.}
\label{fig:gluoncases}
\end{figure*}
To obtain the gluon spectrum for the setup studied in this work, we split it into different contributions to the gluon spectrum according to the longitudinal position where the gluon is emitted in the amplitude and the complex conjugate with respect the hard scattering i.e. if the gluon emission happens before ($bef$) or after ($aft$) the hard scattering.  We recall to the reader that after the hard scattering the gluon can be produced either inside ($in$) or outside ($out$) the QCD medium. Therefore, we have six different possibilities (see Fig. \ref{fig:gluoncases} for a sketch in terms of Feynman diagrams): the direct emissions of the incoming parton (``\emph{bef-bef}") and the outgoing one (``\emph{in-in}",``\emph{in-out}" and ``\emph{out-out}"), and the interferences between both emitters (``\emph{bef-in}" and ``\emph{bef-out}").  The gluon spectrum then reads
\beq
 k^+ \frac{dN_{tot}}{d^3k}= \sum_{i=1}^{6}  k^+ \frac{dN_i}{d^3k}\ . 
\eeq
In the harmonic oscillator approximation the different contributions $k^+ dN_i/d^3k$ are written as
\begin{subequations}
\label{eq:gluoncomp-2}
\beq
\label{eq:befbef-2}
k^+ \frac{dN_{bef-bef}}{d^3k}&=& \frac{\alpha_s\,C_F}{\pi^2} \, \text{Re}\Biggl\{ \int d^2\b \frac{d^2\k'}{(2\pi)^2}\frac{e^{i(\k'-\bkappa)\cdot\b}}{\k'^2}\SO(L^+,0;\b)\Biggr\}\ ,\\
\label{eq:inin-2}
k^+ \frac{dN_{in-in}}{d^3k}&=&\frac{\alpha_s\,C_F}{2\pi^2}\frac{1}{(k^+)^2}\, \text{Re}\Biggl\{\int_0^{L^+} dy'^+\int_0^{y'^+} dy^+\int d^2\b \,e^{-i\bbkappa\cdot\b}\SO(L^+,y'^+;\b)\nonumber\\
&\times&\partial_{\b}\cdot\partial_{\y}\,\KO(y'^+,\b;y^+,\y|k^+)\bigl.\bigr|_{\y=0}
\Biggr\}\ ,\\
\label{eq:inout-2}
k^+ \frac{dN_{in-out}}{d^3k}&=& -\frac{\alpha_s\,C_F}{\pi^2}\frac{1}{k^+}\text{Re}\Biggl\{ 
\int_0^{L^+} dy^+\int d^2\b\,\frac{e^{-i\bbkappa\cdot\b}}{\bbkappa^2}\nonumber\\
&\times&\bbkappa\cdot\partial_\y \KO(L^+,\b;y^+,\y|k^+)\bigl.\bigr|_{\y=0}
\Biggr\}\ ,\\
\label{eq:outout-2}
k^+ \frac{dN_{out-out}}{d^3k}&=&\frac{\alpha_s\,C_F}{\pi^2}\,\frac{1}{\bbkappa^2}\ ,\\
\label{eq:befin-2}
k^+ \frac{dN_{bef-in}}{d^3k}&=& \frac{\alpha_s\,C_F}{\pi^2}\frac{1}{k^+}\text{Re}\Biggl\{\int_0^{L^+}dy^+\int \frac{d^2\k'}{(2\pi)^2}\,d^2\b\,d^2\l\,e^{-i\bbkappa\cdot\b}e^{i\k'\cdot\l}e^{i \delta \k\cdot\l}\nonumber\\
&\times&\SO(L^+,y^+;\b)\frac{\k'\cdot\partial_\b \KO(y^+,\b;0,\l|k^+)}{\k'^2}
\Biggr\}\ ,\\
\label{eq:befout-2}
k^+ \frac{dN_{bef-out}}{d^3k}&=&-2\,\frac{\alpha_s\,C_F}{\pi^2} \text{Re}\Biggl\{ \int \frac{d^2\k'}{(2\pi)^2} \,d^2\b \,d^2\l \frac{\k'\cdot\bbkappa}{\k'^2\,\bbkappa^2} e^{i\k'\cdot\l}e^{-i\bbkappa\cdot\b}e^{i\delta\k\cdot\l} \nonumber\\
&\times&\, \KO (L^+,\b;0,\l|k^+)
\Biggr\}\ ,
\eeq
\end{subequations}
with $\delta \k=\bbkappa-\bkappa$. In the small angle approximation $|\delta\k|\simeq \omega \theta_{qq}$, hence it provides information about the strength of the hard scattering. In the semiclassical method one can allow more general functional forms for the distribution of the colour charges in the QCD medium than the Gaussian white noise implemented here,  see for instance Refs. \cite{Jeon:2005cf,Dumitru:2011ax}. For completeness we write in Appendix \ref{sec:spec-noav} the most general form of the different components of the gluon spectrum without assuming any model for the medium average.  

Notice that when there is no hard scattering, $\theta_{qq}\approx 0$, Eqs. (\ref{eq:gluoncomp-2}) coincide with the results first obtained by Wiedemann (see Eqs. (A-8)-(A-13) of Ref. \cite{Wiedemann:2000tf}) \footnote{Nevertheless,  the analysis of Ref.  \cite{Wiedemann:2000tf} was mainly interested in the case of a ``nascent" parton where just the ``\emph{in-in}", ``\emph{in-out}" and ``\emph{out-out}" were worked out in detail.}. Some of the contributions of the gluon spectrum (\ref{eq:gluoncomp-2}) are already known in the literature. For instance, the ``\emph{in-in}", ``\emph{in-out}" and ``\emph{out-out}" corresponds to gluon emissions due exclusively to the outgoing quark and we identify them together with the BDMPS-Z + vacuum spectrum\footnote{For a complete discussion of the physics of the BDMPS-Z spectrum, we refer to the reader to Refs. \cite{MehtarTani:2012cy,Mehtar-Tani:2013pia}. } \cite{Baier:1998yf,Baier:1996kr,Baier:1996sk,Wiedemann:2000ez,Wiedemann:2000za,Gyulassy:2000fs,Gyulassy:2000er, Zakharov:1996fv,Zakharov:1998sv,Salgado:2003gb,Zakharov:1997uu}.  In the rest of this section we briefly describe the main physical aspects of each contribution to the gluon spectrum. 

\subsection{Direct emissions}
\label{subsec:direm}
Here we describe some of the main properties of the direct emissions of the incoming (``\emph{bef-bef}") and outgoing parton (`\emph{in-in}", ``\emph{in-out}" and ``\emph{out-out}"). We point out the relation between some of the results presented in this section and what it is already known in the literature.

\begin{itemize}
\item [{\bf (i)}] $\textbf{ The ``\emph{bef-bef}" contribution}$:  after performing some of the integrals, Eq. (\ref{eq:befbef-2}) is rewritten as
\beq
\label{eq:befcomp}
k^+ \frac{dN_{bef-bef}}{d^3k}&=&\frac{\alpha_s\,C_F}{\pi^2} \, \text{Re}\Biggl\{ \int d^2\b \frac{d^2\k'}{(2\pi)^2}\frac{e^{i(\k'-\bkappa)\cdot\b}}{\k'^2}\,e^{-\frac{Q_s^2 \b^2}{4}}\Biggr\}\nonumber \\
&=&\frac{\alpha_s\,C_F}{\pi^2}\, \text{Re}\Biggl\{\int \frac{d^2\k'}{(2\pi)^2}\frac{\P(\k'-\bkappa, L^+)}{\k'^2}\,
\Biggr\}\,,
\eeq
where $Q_s^2=\hat{q} L^+$ and we define in Eq. (\ref{eq:befcomp}) the function $\P (\k,\xi)$ as
\beq
\label{eq:Pdef}
\P (\k, \xi)=\frac{4 \pi}{\hat q \xi}\exp\Biggl[-\frac{\k^2}{\hat q \xi}\Biggr]\,.
\eeq
$\P(\k,\xi)$ is a normalized distribution which has a statistical interpretation: it is the probability of having accumulated a certain transverse momentum $\k^2$ while traversing a longitudinal distance $\xi$. As a matter of fact, $\P(\k,\xi)$ accounts for the classical transverse momentum broadening of the gluon after being freed from its parent parton in the medium. 

The medium-induced component Eq. (\ref{eq:befcomp}) can be understood as a two-step process. A gluon with momentum $\k'$ is emitted before of reaching the QCD medium by the incoming parton, the term $1/\k'^2$ in Eq. (\ref{eq:befcomp}), and afterwards the gluon suffers a classical Brownian motion while crossing entirely the medium described by $\P(\k'-\bkappa, L^+)$. Emitted gluons from the initial state get their momenta reshuffled and they acquire in average the maximal amount of accumulated transverse momentum $Q_s=\sqrt{\hat q L^+}$. Moreover,  the multiplicity of gluons associated to the initial state radiation does not change due to medium interactions, instead its angular distribution gets smeared out to large angles. 

\item [{\bf (ii)}] $\textbf{The ``\emph{in-in}" term}$: this corresponds to the medium-induced component of the outgoing parton. Some of the integrals in Eq. (\ref{eq:inin-2}) are easily performed so one can reduce this expression as follows:
\beq
\label{eq:inmed}
\hspace{-1cm}k^+ \frac{dN_{in-in}}{d^3k}&=&  \frac{\alpha_s C_F}{2\pi^2}
\,\frac{1}{(k^+)^2}\,\text{Re}\Biggl\{
2 i k^+\,\int_0^{L^+}dy^+\int\frac{d^2\k'}{(2\pi)^2}\,\P(\k'-\bbkappa,L^+-y^+)\nonumber\\
&\times&\exp\Bigl((1-i)\frac{\k'^2}{2k_f^2}\tan\bigl(\Omega y^+\bigr)\Bigr)
\Biggr\},
\eeq
where $k_f^2=\sqrt{k^+ \hat q}$. Eq. (\ref{eq:inmed}) shows that the medium-induced component can be understood as a two-step process. First, the quantum emission of a gluon with momentum $\k'$  at the time $\tau_f\sim |\Omega|^{-1}$ (the exponential term in the second line of Eq. (\ref{eq:inmed})) and afterwards, the subsequent random walk motion of the gluon along the remaining path through the medium which is described by $\P(\k'-\bbkappa,L^+-y^+)$. The emission spectrum peaks around $k_f$ which corresponds to the amount of  momentum accumulated during its formation time $\tau_f$. The medium-induced component (\ref{eq:inmed}) scales with the length of the medium $L^+$ since gluon emissions can take place along any longitudinal position $0\leq y^+\leq L^+$ inside the medium.

\item [{\bf (iii)}] $\textbf{The ``\emph{in-out}" and ``\emph{out-out}" contributions}$: these terms take into account  gluon radiation when it  takes place either inside the medium or outside it. 

The medium-vacuum interference of the final state radiation is accounted for by the ``\emph{in-out}" contribution, Eq. (\ref{eq:inin-2}). This term can be integrated exactly
\beq
\label{eq:inout-exact}
k^+ \frac{dN_{in-out}}{d^3k}&=&\frac{\alpha_s\,C_F}{\pi k^+}\text{Re}\Biggl\{\int_0^{L^+}dy^+\frac{e^{-i\frac{\bbkappa^2}{2k^+\Omega}\tan\bigl(\Omega(L^+-y^+)\bigr)}}{i\cos^2(\Omega (L^+-y^+))}
\Biggr\}\nonumber\\
&=&-2\frac{\alpha_s\,C_F}{\pi^2}\frac{1}{\bbkappa^2}\text{Re}\Biggl\{
1-\exp\Biggl[-i\frac{\bbkappa^2}{2k^+\Omega}\tan (\Omega L^+)
\Biggr]
\Biggr\}.
\eeq
The ``\emph{in-out}" term becomes important when emissions take place near to the boundary of the medium. The ``\emph{out-out}" component (\ref{eq:outout-2}) takes place completely outside the medium due to  bremsstrahlung of the outgoing parton. 
\end{itemize}

\subsection{Interferences}
\label{subsec:interf}
In the presence of a QCD medium, the information about the colour correlation between both emitters is encoded in the interference terms ``\emph{bef-in}" (\ref{eq:befin-2}) and ``\emph{bef-out}" (\ref{eq:befout-2}). 

The colour correlation between the initial and final state radiation in the presence of a QCD medium is controlled by the path integral $\KO$ of the system formed by the outgoing quark and the gluon radiated by the initial state. Therefore the colour connection between the initial and final state radiation will depend on whether the medium is able to resolve the quark-gluon system or not. This is analogous to the studied case of the $q\bar q$ antenna  immersed in a QCD medium, where the role of $\KO$ is played by the medium decoherence parameter $\Delta_{med}\sim 1-e^{-Q_s^2r_\perp^2}$ \cite{MehtarTani:2010ma,CasalderreySolana:2011rz,MehtarTani:2011gf,MehtarTani:2012cy,MehtarTani:2011tz}, with $r_\perp$  the transverse $q\bar q$ dipole size. However, in the case of the antenna $r_\perp \sim \theta_{q\bar q}\,y^+$ i.e. it  grows linearly with time as it moves along the medium and it depends explicitly on the initial opening angle of the $q\bar q$ pair. In our setup the transverse size of the quark-gluon system depends on the accumulated transverse momentum of the gluon at a certain longitudinal position inside the medium due to its random walk motion; the angle $\theta_{qq}$ determines the initial transverse size of the quark-gluon system when entering in the QCD medium.  The interference pattern between the initial and final state in our setup will depend on the total length of the medium compared with its formation time. We will discuss this detail in length in Sect. \ref{sec:ftan}.

\begin{itemize}
\item  [{\bf (i)}] $\textbf{ The ``\emph{bef-in}" contribution}$: 
the integrals in Eq. (\ref{eq:befin-2}) can be expressed as follows
\beq
\label{eq:befin-comp}\hspace{-1cm}
k^+ \frac{dN_{bef-in}}{d^3k}&=& -2\frac{\alpha_s\,C_F}{\pi^2}\text{Re}\Biggl(i\int_0^{L^+}dy^+\int \frac{d^2\k'}{(2\pi)^2} \frac{\k'\cdot\left(\k'-\delta\k\cos\left(\Omega y^+\right)\right)}{\left(\k'-\delta\k\cos\left(\Omega y^+\right)\right)^2}\nonumber \\ &\times&
\frac{\P(\k'-\bbkappa,L^+-y^+)}{2k^+}
\exp\Biggl\{{(1-i)\frac{\k'^2}{k_f^2}\tan\left(\Omega y^+\right)}\Biggr\}\nonumber
\\
&\times&
\Biggl[1-\exp\Biggr\{i\frac{\left(\k'-\delta\k\cos(\Omega y^+)\right)^2}{2k^+\Omega\sin(\Omega y^+)\cos(\Omega y^+)}
\Biggr\}
\Biggr]\Biggr)\,.
\eeq
The asymptotic behaviour of the latest expression in some particular kinematical regime is determined by how large or small is the argument of the phases. We anticipate that this contribution plays an important role in the incoherent regime $\tau_f\ll L^+$ as we shall see in Sect. \ref{subsec:incohlim}. 

\item [{\bf (ii)}] $\textbf{ The ``\emph{bef-out}" contribution}$: Eq. (\ref{eq:befout-2}) is integrated completely in an analytic manner and it results
\beq
\label{eq:befout-exact}
\hspace{-2cm}k^+ \frac{dN_{bef-out}}{d^3k}&=&2\frac{\alpha_s C_F}{\pi^2}\text{Re}\Biggl\{\int \frac{d^2\k'}{(2\pi)^2}\frac{\k'\cdot\bbkappa}{\k'^2\bbkappa^2}\frac{2\pi i}{k^+\Omega\sin (\Omega L^+)}\nonumber\\
&\times& \exp\Biggl[-\frac{(1-i)}{2k_f^2}\cot(\Omega L^+)\Biggl((\k'+\delta\k)^2+\bbkappa^2-2\frac{\bbkappa\cdot (\k'+\delta\k)}{\cos (\Omega L^+)}\Biggr)
\Biggr]
\Biggr\}\ ,\nonumber\\
&=&-2\frac{\alpha_s C_F}{\pi^2}
\text{Re}\Biggl\{
\frac{\bbkappa\cdot\bigl(\bbkappa-\delta\k \cos (\Omega L^+)\bigr)}{\bbkappa^2 \bigl(\bbkappa-\delta\k \cos (\Omega L^+)\bigr)^2}
\exp\Biggl[(1-i)\frac{\bbkappa^2}{2k_f^2}\tan(\Omega L^+)\Biggr]\nonumber\\
&\times&\Biggl(1-\exp\Biggl[i\frac{\bigl(\bbkappa-\delta\k\cos (\Omega L^+)\bigr)^2}{2k^+\Omega\,\sin (\Omega L^+)\cos (\Omega L^+)}
\Biggr]
\Biggr)
\Biggr\}\ . 
\eeq
The ``\emph{bef-out}" contribution is the responsible for the decoherence of the vacuum radiation. This is clearly seen when taking the limit  $|\Omega| L^+\ll 1$ in Eq. (\ref{eq:befout-exact}) since in this limit the ``\emph{bef-out}" term is proportional to $\bkappa\cdot\bbkappa/(\bkappa^2\bbkappa^2)$ i.e. the vacuum interference term cf. the last term in Eq. (\ref{eq:vacspec}). 
\end{itemize}

\subsection{Recovering the vacuum coherence pattern}
\label{sec:vaclim}

Before starting the analysis of the physical scales involved the gluon spectrum, we show here that, in the absence of a QCD medium,  Eqs. (\ref{eq:gluoncomp-2}) reduce to the genuine vacuum gluon spectrum.  This is easily obtained by taking the limit $\hat q \to 0$ in each contribution in Eqs. (\ref{eq:gluoncomp-2}), resulting in
\begin{subequations}
\label{eq:vaclimit}
\beq
k^+ \frac{dN_{bef-bef}}{d^3k}&=&\frac{\alpha_s C_F}{\pi^2}\frac{1}{\bkappa^2}\,, \\
k^+ \frac{dN_{in-in}}{d^3k}&=&2 \frac{\alpha_s C_F}{\pi^2}\frac{1}{\bbkappa^2}\Biggr[1-\cos\Biggl(\frac{\bbkappa^2}{2k^+}L^+\Biggr)\Biggl]\,, \\
k^+ \frac{dN_{in-out}}{d^3k}&=&-2 \frac{\alpha_s C_F}{\pi^2}\frac{1}{\bbkappa^2}\Biggr[1-\cos\Biggl(\frac{\bbkappa^2}{2k^+}L^+\Biggr)\Biggl] \,,\\
k^+ \frac{dN_{out-out}}{d^3k}&=&\frac{\alpha_s\,C_F}{\pi^2}\,\frac{1}{\bbkappa^2}\,,\\
k^+ \frac{dN_{bef-in}}{d^3k}&=&2\frac{\alpha_s\,C_F}{\pi^2}\,\frac{\bkappa\cdot\bbkappa}{\bkappa^2\,\bbkappa^2}\Biggl[\cos\Biggl(\frac{\bbkappa^2}{2k^+}L^+\Biggr)-1\Biggr]\,,
\\
k^+ \frac{dN_{bef-out}}{d^3k}&=&-2\frac{\alpha_s\,C_F}{\pi^2}\,\frac{\bkappa\cdot\bbkappa}{\bkappa^2\,\bbkappa^2}\cos\Biggl(\frac{\bbkappa^2}{2k^+}L^+\Biggr)\,.
\eeq
\end{subequations}
When added, the gluon spectrum for $\hat{q}\to 0$ reads
\beq
\label{eq:gluvac}
 k^+ \frac{dN_{tot}}{d^3k}\Biggl.\Biggr|_{\hat q \to 0}&=& \sum_{i=1}^{6}  k^+ \frac{dN_i}{d^3k}\Biggl.\Biggr|_{\hat q \to 0}\nonumber \\
 &=&\frac{\alpha_s C_F}{\pi^2}\Biggl(\frac{1}{\bkappa^2}+\frac{1}{\bbkappa^2}-2\frac{\bkappa\cdot\bbkappa}{\bkappa^2\,\bbkappa^2}\Biggr)\nonumber\\
 &=&\frac{\alpha_s C_F}{2\pi^2}\frac{p\cdot\bar p}{(p\cdot k)(\bar p\cdot k)}\ ,
\eeq
which we recognise as the vacuum gluon spectrum (see Eq. (\ref{eq:vacspec}) in Appendix \ref{sec:vac-spec}).  In the third line we make use the definitions of $\bkappa=\k-x\p$ (likewise for $\bbkappa$). 

The inclusive gluon spectrum (\ref{eq:gluvac}) presents both soft and collinear divergences.  Apparently, interference terms may look like non-ladder Feynman diagrams which would spoil the usual collinear factorisation. Nevertheless, it is possible to separate  gluon radiation off the incoming and outgoing partons by accounting the interference terms and, thus, to provide a probabilistic interpretation. Angular ordering is precisely what allows such probabilistic interpretation. To see it more clearly, let us consider the total number of gluons by integrating Eq. (\ref{eq:gluvac}) over the transverse momentum and the frequency of the gluon. For simplicity and without losing the generality of the result, let us consider this average  along the longitudinal axis of the outgoing quark. The result of this simple exercise, considering the contribution from the incoming and outgoing quarks, is \cite{Dokshitzer:1991wu}
\beq
\label{eq:incvacspec}
N_g &=&  \frac{\alpha_s \,C_F}{\pi}\,\frac{1}{2}\log^2 \Biggl(\frac{Q^2}{Q_0^2}\Biggr),
\eeq
where $Q_0$ is a momentum cut-off to cure the collinear and soft divergences of the emitted gluon and $Q^2=(p^+\theta_{qq})^2$ is the virtuality of the photon associated to the hard scattering. The factor $1/2$ is a consequence of angular ordering (the other one coming from taking just one emitter): if the angular restriction is not considered, the number of incoherent gluon emissions will be twice the coherent one. Eq. (\ref{eq:incvacspec}) also indicates that part of these logarithms must be resummed into the fragmentation function of the final state. Similarly,  half the result Eq. (\ref{eq:incvacspec}) follows from the respective integrations with respect to the initial state and  the logarithms are resummed in the parton density functions (PDFs). The generalisation of this result when describing multiple gluon emissions in a parton cascade leads to a modification of the DGLAP evolution equation which is known as Modified Leading Logarithmic Approximation (MLLA) \cite{Dokshitzer:1991wu,Bassetto:1984ik,Ellis:1991qj}. 

\section{Formation time analysis}
\label{sec:ftan}

In order to understand and gain physical intuition of the gluon spectrum (\ref{eq:gluoncomp-2}), we perform in this section some analytical approximations by considering the relevant scales of the problem\footnote{In this Section and in the following one, we will use the gluon energy $\omega$ and the gluon $k^+$  light-cone coordinate as equivalent, the difference among these variables at high energies and in the small angle approximation is a factor $\sqrt{2}$ that affects neither the qualitative parametric employed arguments  nor the calculations to the accuracy that they are performed.}. First, we notice that the oscillatory behaviour of the argument of the phases (\ref{eq:A-Bcoeff}) in the path integral $\KO$ (\ref{eq:Kosc})  will depend on the ratio $\Delta\tau/\tau_f$ ($\tau_f=\sqrt{k^+/\hat{q}}\sim|\Omega|^{-1}$). $\tau_f$ is to be understood as the time a gluon fluctuation takes to colour-decohere from the quark-gluon system state. It should not be confused with the typical formation time in vacuum $\sim \omega/\k^2$ since its magnitude depends on the transport properties of the medium via $\hat{q}$.

If $\Delta\tau\sim L^+$, say the longitudinal distance is on the order of the longitudinal size of the medium, the ratio $L^+/\tau_f$ separates the spectrum into two regimes: (i) the coherent regime when  $L^+\ll\tau_f$ and (ii) the incoherent regime when $L^+\gg\tau_f$\footnote{In the language used in previous literature \cite{Salgado:2003gb}, these two limits corresponds to the case when $\omega \gg \omega_c$ (coherent regime) and $\omega \ll \omega_c$ (incoherent regime), with $\omega_c=\hat q (L^+)^2/2$.}. 
In what follows we study separately each of these limits on the gluon spectrum (\ref{eq:gluoncomp-2}) and describe its main properties by considering some analytical approximations. 
\subsection{Coherent limit: $L^+\ll\tau_f$}
\label{subsec:cohlim}
This limit corresponds to the physical situation when the emitted gluon remains coherent during all the time while crossing the QCD medium.
In this case, one enters the deep Landau-Migdal-Pomeranchuck (LPM) regime where the medium acts as a unique scattering center so the squared effective momentum transfer is  $\sim Q_s^2=\hat{q}L^+$ that is constant. As a consequence, $dN_{in-in}$, $dN_{bef-in}$ and $dN_{in-out}$ are suppressed. Therefore, the total spectrum is given by the sum of three components 
\beq
 k^+ \frac{dN_{tot}}{d^3k}\Biggl.\Biggr|_{\tau_f\gg L^+}= k^+ \frac{dN_{bef-bef}}{d^3k}+k^+ \frac{dN_{bef-out}}{d^3k}+k^+ \frac{dN_{out-out}}{d^3k}\,,
\eeq
where it must be understood that the  ``\emph{bef-bef}", ``\emph{bef-out}" and ``\emph{out-out}" contributions in the right hand side of this expression are evaluated for $L^+\ll\tau_f$. 
So after integrating exactly the ``\emph{bef-bef}" term (\ref{eq:befcomp}), in this limit all the three non-vanishing components of the gluon spectrum read
\begin{subequations}
\label{eq:spec-coh}
\beq
\label{eq:befcoh}
k^+ \frac{dN_{bef-bef}}{d^3k}&=&\frac{\alpha_s\,C_F}{\pi^2}\, \text{Re}\Biggl\{\int \frac{d^2\k'}{(2\pi)^2}\frac{\P(\k'-\bkappa, L^+)}{\k'^2}\,
\Biggr\}\,\nonumber \\
&=&\frac{\alpha_s C_F}{\pi^2}\,\frac{e^{-\bkappa^2/Q_s^2}}{Q_s^2}\nonumber \\
&\times& \text{Re}\left\{\log\Biggl(\frac{Q_s^2}{Q_0^2}\Biggr)-\Gamma\Biggl(0,-\frac{\bkappa^2}{Q_s^2}\Biggr)-\gamma_E-\log\Biggl(-\frac{\bkappa^2}{Q_s^2}\Biggr)
\right\},\\
\label{eq:outcoh}
k^+ \frac{dN_{out-out}}{d^3k}&=&\frac{\alpha_s\,C_F}{\pi^2}\,\frac{1}{\bbkappa^2}\ ,\\
\label{eq:intcoh}
k^+ \frac{dN_{bef-out}}{d^3k}&=&-2\frac{\alpha_s\,C_F}{\pi^2}\,\frac{\bkappa\cdot\bbkappa}{\bkappa^2\bbkappa^2}\Bigl(1-e^{-\bkappa^2/Q_s^2}\Bigr).
\eeq
\end{subequations}
In Eq. (\ref{eq:befcoh}) $\Gamma[a,z]$ is the incomplete Gamma function, $\gamma_E$ is Euler's constant and  $Q_0$ is an infrared cut-off which cures the collinear divergence associated to the initial state.  Eqs. (\ref{eq:spec-coh}) shows that the gluon spectrum depends on the scales associated to the hard scattering $\delta\k$ and the typical momentum transfer of the medium  $Q_s$.  In the rest of this subsection we focus on how the expressions (\ref{eq:spec-coh}) behave while varying $\k$ for (i) out-of-cone emissions in the regime $Q_s^2\gg \delta\k^2$ i.e. radiation outside the cone delimited by $\theta_{qq}$, $\theta>\theta_{qq}$, or (ii) in-cone emissions, $\theta\ll \theta_{qq}$, in the regime $\delta\k^2\gg Q_s^2$.

\subsubsection{Out-of-cone emissions}
\label{sub:largeangle-coh}
When $\delta\k^2\ll Q_s^2$  the hard scattering does not affect much the trajectory of the highly energetic parton. In the high energy limit when $\delta\k\equiv 0$ then $\bkappa\sim\bbkappa\equiv\k$ and Eqs. (\ref{eq:spec-coh}) coincide exactly with the well-known result derived by Kovchegov and Mueller, see  Eqs. (59-61) in Ref. \cite{Kovchegov:1998bi}\footnote{We thank Prof. Mueller for pointing us out the relation between the results of Ref. \cite{Kovchegov:1998bi} and our work. In order to make a straightforward comparison between both results,  the cut-off  for curing the collinear divergence associated to the initial state is written in this work in momentum space while in Ref. \cite{Kovchegov:1998bi} it is done in coordinate space. Both prescriptions can be easily mapped one into each other.}. The high-energy limit of  gluon production in the dilute-dense regime within the Color Glass Condensate framework and related approaches, has also been studied by different authors \cite{Kovchegov:2001sc,Kovchegov:1998bi,Dumitru:2001ux,Kopeliovich:1998nw,Blaizot:2004wu,JalilianMarian:2005jf} in the past. 

When $\delta\k^2,\k^2\ll Q_s^2$ the gluon spectrum (\ref{eq:spec-coh}) reduces to
\beq
\label{eq:coh-k<Qs}
k^+ \frac{dN_{tot}}{d^3k}\Biggl.\Biggr|_{\delta\k,\k\ll Q_s}&\simeq& \frac{\alpha_s C_F}{\pi^2}\,\Biggl(\frac{1}{\bbkappa^2}+ \text{Re}\Biggl\{\int \frac{d^2\k'}{(2\pi)^2}\frac{\P(\k'-\bkappa, L^+)}{\k'^2}\,
\Biggr\}\,\Biggr)\,\nonumber\\
&=& \frac{\alpha_s C_F}{\pi^2}\,\Biggl(\frac{1}{\bbkappa^2}+ \frac{e^{-\bkappa^2/Q_s^2}}{Q_s^2}\log\Biggl(\frac{Q_s^2}{Q_0^2}
\Biggr)
\Biggr)\ .
\eeq
The first term inside the brackets comes from the ``\emph{out-out}" contribution while the last one corresponds to the leading term in the ``\emph{bef-bef}" contribution, Eq. (\ref{eq:befcoh}). This result can be easily understood: the first term in Eq. (\ref{eq:coh-k<Qs}) takes into account those gluons in the incoming parton's wavefunction with $\k^2\ll Q_s^2$ that are emitted and get broadened i.e. the momentum transfer from the medium opens the phase space for large angle emissions. The second term inside the brackets in Eq. (\ref{eq:coh-k<Qs}) corresponds to the parton undergoing hard scattering that leaves the medium while remaining off-shell. In rebuilding its gluon cloud, further gluon emission by bremsstrahlung is required and these emitted gluons do not experience transverse broadening. 

When $\delta\k^2\ll Q_s^2\ll\k^2$, all the vacuum bremsstrahlung terms cancel when summed together, and the only remaining term is the first one inside the brackets in Eq. (\ref{eq:befcoh}). This term falls off exponentially and thus becomes negligible. This observation implies that the exponential factor $e^{-\k^2/Q_s^2}$ sets up a maximal angle for gluon emissions given by $\theta_{max}\sim Q_s/\omega$. In other words, the vacuum coherence pattern at large angles $\theta>\theta_{max}$ is reestablished. On the other hand, and as a consequence of the interaction with the medium there is a suppression of interferences for soft gluons with $\delta\k^2\ll\k^2\ll Q_s^2$. The lost of vacuum coherence opens the phase space for large angular emissions beyond $\theta=\theta_{qq}$, i.e., there is antiangular ordering in the interval $\theta_{qq}\leq\theta\leq\theta_{max}$. This behaviour was also observed also in the $q\bar q$ antenna immersed in a QCD medium \cite{MehtarTani:2010ma,MehtarTani:2011gf,MehtarTani:2012cy}.

\subsubsection{In-cone emissions}
\label{sub:smallangle-coh}
 If we consider that $|\delta\k|\simeq\omega\theta_{qq}$ is the maximal scale, the spectrum (\ref{eq:spec-coh}) reduces to
\beq
\label{eq:coh-k>Qs}
k^+ \frac{dN_{tot}}{d^3k}\Biggl.\Biggr|_{\delta\k\gg Q_s}&\simeq& \frac{\alpha_s C_F}{\pi^2}\, \frac{e^{-\bkappa^2/Q_s^2}}{Q_s^2}\log\Bigl(\frac{Q_s^2}{Q_0^2}\Bigr)+\frac{\alpha_s C_F}{\pi^2}\, \Biggl(\frac{1}{\bkappa^2}-2\frac{\bkappa\cdot\bbkappa}{\bkappa^2\bbkappa^2}+\frac{1}{\bbkappa^2}\Biggr)\nonumber\\
&\simeq& \frac{\alpha_s C_F}{\pi^2}\, \Biggl(\frac{1}{\bkappa^2}-2\frac{\bkappa\cdot\bbkappa}{\bkappa^2\bbkappa^2}+\frac{1}{\bbkappa^2}\Biggr)\,.
\eeq
The second line of the latest expression follows directly since within this limit $|\bkappa|=|\bbkappa-\delta\k|\gg Q_s$ so the exponential term is negligible. The remaining terms are easily identified with the vacuum gluon spectrum, Eq. (\ref{eq:vacspec}). Therefore, inside the cone of radiation with opening angle $\theta,\theta_{max}<\theta_{qq}$ the medium does not destroy the vacuum coherence pattern between the initial and final state radiation. This follows from the fact that the hard scattering is so hard that the medium cannot modify the specific vacuum features of radiation in this setup. It shows that for large enough virtuality exchanges in the hard scattering, the vacuum structure of collinear divergencies and angular ordering is recovered. Note that we are working in the limit of large formation times in which emissions off the outgoing radiating parton happen outside the medium.

\subsection{Incoherent limit: $L^+\gg\tau_f$}
\label{subsec:incohlim}

In this limit gluons are allowed to be produced anywhere along the medium length. So in order to get the correct leading order behaviour of each of the contributions within this regime, it is necessary to understand if the gluon is produced either at early or late times after its creation point inside the medium. In the gluon spectrum (\ref{eq:gluoncomp-2}) this detail is relevant for the ``\emph{in-in}" (\ref{eq:inin-2}) and ``\emph{in-out}" (\ref{eq:inout-2}) contributions as well as the interference terms ``\emph{bef-in}" (\ref{eq:befin-2}) and ``\emph{bef-out}" (\ref{eq:befout-2}) respectively. The ``\emph{bef-bef}"  and ``\emph{out-out}" terms given by Eqs. (\ref{eq:befcoh}) and (\ref{eq:outcoh}) respectively, have the same analytical form as in the coherent limit.

 In the rest of this section, we work out some approximations which allow us to obtain the correct leading order behaviour of the contributions in this regime.  We conclude this section by summarizing our main findings for the angular distribution of the gluon spectrum in the incoherent limit. 
 
\begin{itemize}
\item [{\bf (i)}] $\textbf{The ``\emph{in-in}" contribution:}$ the oscillatory behaviour of the tangent in (\ref{eq:inmed}) depends on the relation between the longitudinal position where the gluon is emitted and the typical formation time in the medium i.e. $|\Omega| y^+\sim y^+/\tau_f$. Depending on the value of this ratio $y^+/\tau_f$ we have two different cases: either the emissions take place long time after the creation point of the medium, $ L^+\gtrsim y^+\gg \tau_f$, or emissions happen at early times , $ L^+\gg \tau_f\gg y^+$. So it is possible to extract the correct leading order term of Eq. (\ref{eq:inmed}) by analyzing each one of these two cases independently. The details of such calculation were already worked out by some of us in the case of the $q\bar q$ antenna spectrum immersed in a QCD medium,  see Sect. 5 of Ref. \cite{MehtarTani:2012cy}\footnote{For completeness we present the derivation of Eq. (\ref{eq:inin-lead}) in Appendix \ref{sec:ininapp}.}. It turns out that the leading term of the ``\emph{in-in}" contribution in the incoherent regime reads as
\beq
\label{eq:inin-lead}
k^+ \frac{dN_{in-in}}{d^3k}\Biggl.\Biggr|_{\tau_f\ll L^+}&=&\frac{\alpha_s\,C_F}{k^+ \pi^2}\int_{0}^{L^+}dy^+\int \frac{d^2\k}{(2\pi)^2}\P(\k-\bbkappa,L^+-y^+)\nonumber \\
&\times& e^{-\frac{\k^2}{k_f^2}}\sin\Biggl(\frac{\k^2}{k_f^2}\Biggr)
\,.
\eeq
As we discussed previously, medium-induced gluon radiation in Eq. (\ref{eq:inin-lead}) can be understood as a two-step process. The terms in the second line of Eq. (\ref{eq:inin-lead}) take into account the quantum emission of a gluon with momentum $\k$ at the time $y^+$ with a momentum distribution centered around the preferred value $k_f$, while the function $\P(\k-\bbkappa, L^+-y^+)$ describes the classical Brownian motion that the gluon experiences  until the edge of the medium once freed from its parent parton. This component is suppressed for $\k^2 > k_f^2$. Nevertheless, the transverse momentum of the gluon gets reshuffled, with the saturation scale of the medium $Q_s$ as typical scale,  due to the broadening at the end of its trajectory. Another important property of the ``\emph{in-in}" contribution (\ref{eq:inin-lead}) is its scaling with the medium length $L^+$ which is a consequence of the probability of the projectile to emit at any point during its trajectory along the medium layer.  The subleading terms not shown in Eq. (\ref{eq:inin-lead})  are enhanced neither by a logarithmic divergence, see Appendix \ref{sec:ininapp}, nor by the medium length, so they can be safely neglected at this level of our approximations. 

\item [{\bf (ii)}] $\textbf{The ``\emph{in-out}" contribution:}$ in the incoherent limit $|\Omega| L^+\gg 1$ one approximates $\tan (\Omega L^+)\approx -i$ in Eq. (\ref{eq:inout-exact}). Afterwards, it is straightforward to get the leading term of the ``\emph{in-out}" contribution which reads
\beq
\label{eq:inout-incoh}
k^+ \frac{dN_{in-out}}{d^3k}\Biggl.\Biggr|_{\tau_f\ll L^+}&=&-2\frac{\alpha_s\,C_F}{\pi^2}\frac{1}{\bbkappa^2}\Biggl[1-e^{-\frac{\k^2}{k_f^2}}\cos\Biggl(\frac{\k^2}{k_f^2}\Biggr)
\Biggr]\,.
\eeq

\item [{\bf (iii)}] $\textbf{The ``\emph{bef-in}" contribution:}$ as in the case of the ``\emph{in-in}" contribution, here it is important to consider if the gluon is emitted either at early times when $0\leq \ y^+\ll\tau_f$ or at late times when $\tau_f\ll y^+\lesssim L^+$. 

For late-time emissions, the leading order term of Eq. (\ref{eq:befin-comp}) scales like $e^{-\frac{y^+}{\tau_f}}$. Therefore it gets effectively suppressed 
\beq
k^+ \frac{dN_{bef-in}}{d^3k}\Biggl.\Biggr|_{\tau_f\ll y^+\lesssim L^+}\approx 0. 
\eeq
At early times $y^+\ll\tau_f$ one can neglect the last term in the third line of Eq. (\ref{eq:befin-comp}) since it oscillates very fast. Additionally for short formation times $\P(\k-\bbkappa,L^+-y^+)\simeq \P(\k-\bbkappa,L^+)$. So for short formation times both approximations let us to rewrite Eq. (\ref{eq:befin-comp}) as
\beq
\label{eq:befin-incoh-1}
k^+ \frac{dN_{bef-in}}{d^3k}\Biggl.\Biggr|_{0\lesssim y^+\ll\tau_f}&\simeq&-2 \frac{\alpha_s\,C_F}{\pi^2}\text{Re}\Biggl\{\int_{0}^{\tau_f}dy^+\int \frac{d^2\k'}{(2\pi)^2} 
\frac{\k'\cdot(\k'-\delta\k)}{(\k'-\delta\k)^2}
\nonumber\\
&\times& \,\P(\k'-\bbkappa,L^+)\exp\Biggl[{\frac{(1-i)}{2k_f^2}\k'^2\tan\left(\Omega y^+\right)}
\Biggr]
\Biggr\}\,.
\eeq
Next, we can integrate over $y^+$ exactly. However, as we explain in Appendix \ref{sec:ininapp} (see discussion after Eq.(\ref{eq:exp-integral}), the exact integral has terms which are not proportional to $L^+$ so we can drop them out. Thus, we concentrate in the case where $\k'^2>k_f^2$ where the integral over $y^+$ can be approximated as it is described in Eq. (\ref{eq:exp-integral-2}) in Appendix \ref{sec:ininapp}. Finally Eq. (\ref{eq:befin-incoh-1}) is given by
\beq
\label{eq:befin-incoh-2}
\hspace{-0.5cm}k^+ \frac{dN_{bef-in}}{d^3k}\Biggl.\Biggr|_{0\lesssim y^+\ll\tau_f}&\simeq&-2 \frac{\alpha_s\,C_F}{\pi^2}\text{Re}\Biggl\{\int\frac{d^2\k'}{(2\pi)^2}\frac{\k'\cdot(\k'-\delta\k)}{\k'^2\left(\k'-\delta\k\right)^2}\P(\k'-\bbkappa,L^+)
\Biggr\}\,,
\eeq
which is valid for $\k'^2>k_f^2$. Notice the similarity of this expression and the exact ``\emph{bef-bef}" contribution (\ref{eq:befcomp}). Therefore, this term looks like a hard vacuum-like interference component at $\k'^2>k_f^2$ that is generated at early times in the medium and subsequently follows a random walk process. 

\item  [{\bf (iv)}] $\textbf{The ``\emph{bef-out}" contribution:}$ while taking the incoherent limit of Eq. (\ref{eq:befout-exact}), one finds that the ``\emph{bef-out}" contribution   scales as $\sim e^{-|\Omega| L^+}$ so it is exponentially suppressed in this regime. Thus,  we can safely neglect this contribution in this regime.  
\end{itemize}
All the analyses carried out above indicate  that in the incoherent regime,  the non-vanishing contributions to the gluon spectrum (\ref{eq:gluoncomp-2}) are the direct emissions of the initial and final state radiation and the early-time emission of the ``\emph{bef-in}" interference term:
\beq
\label{eq:totspec-incoh}
k^+ \frac{dN_{tot}}{d^3k}\Biggl.\Biggr|_{\tau_f\ll L^+}&=& k^+ \frac{dN_{bef-bef}}{d^3k}+k^+ \frac{dN_{in-in}}{d^3k}+k^+ \frac{dN_{in-out}}{d^3k}+k^+ \frac{dN_{out-out}}{d^3k}\nonumber \\&+&k^+\frac{dN_{bef-in}}{d^3k}\,,
\eeq
where the terms 
``\emph{bef-bef}", 
``\emph{in-in}", 
``\emph{in-out}",
``\emph{out-out}" and  ``\emph{bef-in}"  contributions are given by Eqs. (\ref{eq:befcoh}), (\ref{eq:inin-lead}), (\ref{eq:inout-incoh}), (\ref{eq:outcoh}) and (\ref{eq:befin-incoh-2}) respectively. 

In this regime, we also have a gradual decoherence in the region between $\delta\k^2\ll\k^2\ll Q^2_s$. Besides, the interference contribution (\ref{eq:befin-incoh-2}) restores the vacuum coherence pattern above $Q_s$. This can be easily understood since, for $\k^2>Q_s^2$, the effects of the broadening do not affect any more the final state transverse momenta of the emitted gluon, and thus $\P(\k-\bbkappa,L^+)\left.\right|_{\k^2>Q_s^2}\simeq (2\pi)^2\delta^{2}(\k-\bbkappa)$. As a result, for $\k^2>Q_s^2$ the ``\emph{bef-in}" term (\ref{eq:befin-incoh-2}) becomes $\sim -2\bbkappa\cdot\bkappa/(\bbkappa^2\bkappa^2)$ which is precisely the hard vacuum-like interference term that cancels large angle emissions. Therefore there is again a maximal angle $\theta_{max}$ which sets up an upper bound for large angle emissions. 

The physical picture of this regime turns out to have a remarkable intuitive interpretation. We have two emitters, the initial and final state radiators with a non-negligible interference term. For the initial state radiation the medium reshuffles the momenta of the emitted collinear gluon up to $\k^2\lesssim Q_s^2$ and there is a maximal angle of emission $\theta_{max}\sim Q_s/\omega$ caused by the restoration of the vacuum coherence due to the vacuum-like interference term (\ref{eq:befin-incoh-2}). The contribution associated to the final state is composed by the genuine medium-induced radiation BDMPS-Z + vacuum emission. Both terms emit gluons up to $\theta_{max}= Q_s/\omega$. We clarify that BDMPS-like emissions have an upper bound for large angle emissions since the broadening of the gluon by the medium sets $Q_s$ as the hardest scale of the process. Bremsstrahlung-like radiation outside the medium radiates coherently since it is bounded at large angles due to the non-vanishing interference term  (\ref{eq:befin-incoh-2}) as we discussed previously. 

So far we have focused on the discussion of the medium modifications of the angular distributions of gluon emissions. In both regimes we observe the loss of vacuum coherence between both emitters depending on the interplay between two scales, the opening angle $\delta\k^2\sim (\omega \theta_{qq})^2$ and the maximum momentum transferred by the medium $Q_s$. As a consequence, the total number of observed gluons will be modified according with the dominant scale. Hence, there is a hint for a possible modification to the quantum evolution of the parton densities and fragmentation functions inside nuclei. This aspect deserves a more general discussion that we make in more detail in the next Section. 
 
\section{Tracing modifications of quantum evolution}
\label{sec:discussion}

In the previous section we discussed the coherent and incoherent limits of the gluon spectrum (\ref{eq:gluoncomp-2}) by considering formation time arguments. In both regimes, we observe either a partial or a complete loss of vacuum interference between the initial and final state radiation due to the medium interactions which opens the phase space for large angle gluon emissions. The decoherence between the two emitters in both regimes depends strongly on the competition between two main scales: the opening angle due to the hard scattering $\delta\k^2\sim (\omega \theta_{qq})^2$ and the maximum momentum transferred by the medium $Q_s^2=\hat{q}L^+$. In this Section, we compute the dominant logarithmic contribution to the total number of gluons observed at certain opening angle which carry information about the dominant scale. Our computations are performed in a similar manner as in the vacuum case Sect. (\ref{sec:vaclim}).

We proceed to examine the different contributions to the single inclusive spectrum and keep those that contain logarithmic contributions. Notice that when taking into account the broadening of the emitted gluon in the medium (given by function $\P (\k,\xi)$, Eq. (\ref{eq:Pdef})), it does not change the total number of produced gluons but simply reshuffles their transverse momentum distribution\footnote{This is a consequence of the normalization of the function $\P (\k,\xi)$.}. Thus:
\begin{itemize}

\item The ``\emph{bef-bef}" contribution, Eq. (\ref{eq:befcomp}), contains a collinear vacuum contribution from the incoming parton.

\item The ``\emph{in-in}"  and ``\emph{in-out}" contributions, Eqs. (\ref{eq:inmed}) and (\ref{eq:inout-exact}), are BDMPS-Z contributions that contain neither infrared not collinear divergencies, and therefore they do not contribute to the multiplicity to logarithmic accuracy.

\item The ``\emph{out-out}" contribution, Eq. (\ref{eq:outout-2}), gives a purely vacuum contribution from the outgoing parton.

\item The ``\emph{bef-out}" contribution, Eq. (\ref{eq:befout-exact}), provides a vacuum interference term for large frequencies $\omega>\omega_c$, as we discuss in Sections \ref{subsec:interf} and \ref{subsec:cohlim}.

\item The ``\emph{bef-in}" contribution, Eq. (\ref{eq:befin-comp}), contains a  (non-logarithmically behaving) medium-induced piece for $\k^2<k_f^2$, and a vacuum interference term for $\k^2>k_f^2$, for small frequencies $\omega<\omega_c$, as we pointed out in the discussion in Sect. \ref{subsec:incohlim} (see the paragraphs after Eqs. (\ref{eq:befin-incoh-1}) and (\ref{eq:befin-incoh-2})).

\end{itemize}

If we consider for instance the following hierarchy of scales: $|\delta \k|\sim \omega \theta_{qq} < k_f=(\omega \hat q)^{1/4} < \omega$, we can  complete the vacuum interference in the ``\emph{bef-in}" contribution by adding a  piece that is integrated in $|\k|$ from $|\delta \k|$ to $k_f$. In this way we get the full vacuum Eq. (\ref{eq:incvacspec}) minus a medium contribution. Because of the minus sign in Eq. (\ref{eq:befin-comp}), this medium contribution is positive and it reads, to double logarithmic accuracy as follows
\beq
\label{eq:ngmedredux}
N_g^{med}&\simeq&
\frac{2 \alpha_s C_F}{\pi^2}\int_{\omega_{min}}^{\omega_{max}} \frac{d\omega}{\omega} \int_{|\delta \k|}^{k_f} \frac{d^2\k}{\k^2}\nonumber \\
&= &\frac{1}{6}\frac{\alpha_s C_F}{\pi} \left[  \log^2\left(\frac{\hat q}{\theta_{qq}^4 \omega_{min}^3} \right)-\log^2 \left( \frac{\hat q}{ \theta_{qq}^4 \omega_{max}^3}  \right)    \right]\nonumber \\
&\simeq &\frac{1}{6}\frac{\alpha_s C_F}{\pi} \left[  \log^2\left(\frac{1}{\theta_{qq}^4} \right)-\log^2 \left( \frac{\theta_c^4}{ \theta_{qq}^4}  \right)    \right]\nonumber \\
&=& \frac{8}{3}\frac{\alpha_s C_F}{\pi}\,  \log\left(\frac{\theta_c}{\theta_{qq}^2} \right) \log \left( \frac{1}{ \theta_{c}}  \right) ,
\eeq 
where $\omega_{max}=\omega_c=\hat{q}L^2/2$ (thus $\theta_{qq}< \theta_c\ll 1$), $\omega_{min}=\hat{q}^{1/3}$ from the assumed hierarchy of scales\footnote{Note that $|\delta \k| >Q_0$, with $Q_0$ some infrared cut-off, implies $\omega>Q_0/\theta_{qq}$. Here we assume $Q_0/\theta_{qq}<\hat q^{1/3}$.}, and $\theta_c=2/\sqrt{\hat q L^3}$ is the characteristic BDMPS-Z emission angle, see e.g. \cite{CasalderreySolana:2007zz}. 
In this expression we see clearly the competition between the scales of the emitters and of the medium. From Eq. (\ref{eq:ngmedredux}) we observe that the medium contribution is completely suppressed for large opening angles $\theta_{qq}$ as expected. When increasing medium length or density, the characteristic BDMPS-Z angle decreases and the medium contribution increases.

In the opposite case, $\theta_c<\theta_{qq}\ll 1$, the integration over $\omega$ is bounded from above by $\omega_{max}=\hat q^{1/3}/\theta_{qq}^{4/3}<\omega_c $ and we get 
\beq
\label{eq:ngmedredux2}
N_g^{med}&\simeq&\frac{8}{3}\frac{\alpha_s C_F}{\pi}\,  \log^2\left(\frac{1}{\theta_{qq}} \right) \, .
\eeq

Our results (\ref{eq:ngmedredux}) and (\ref{eq:ngmedredux2}) are done under a given assumed hierarchy of scales $|\delta\k| < k_f < \omega$. These expressions can be easily modified for other situations, for instance when $\omega_{min}=Q_0/\theta_{qq}>\hat q^{1/3}$,  or $\omega_{max}=p^+< \omega_c$. In these cases, additional competitions to the existent ones between the medium parameters and the opening angle i.e. those originated from the presence of the infrared cut-off or the energy, start to play a role. Nevertheless, the existence of double logarithmic contributions remains.

These additional, potentially large logarithms that we find on top of the vacuum ones, Eq. (\ref{eq:incvacspec}), suggest an alteration of the usual perturbative DGLAP \cite{Dokshitzer:1977sg,Gribov:1972ri,Gribov:1972rt,Altarelli:1977zs} evolution due to a QCD medium of finite size. This aspect indeed deserves further studies that we leave for a future publication. 

\section{Conclusions}
\label{sec:conclusions}

In this work we investigate the medium modifications to the interference pattern between  initial and final state radiation.  We investigate gluon radiation off an `asymptotic' parton which suffers a hard scattering and afterwards crosses a dense QCD medium of finite length. We calculate the general form of the gluon spectrum and then we formulate it within the harmonic oscillator approximation. In this approximation we are able to get an analytical treatment and understanding of the different components of the gluon spectrum. Let us note that, while we consider a colourless $t$-channel exchange as hard scattering, the results for soft gluons do not change from the ones for a coloured exchange, as it is well known in the vacuum  \cite{Dokshitzer:1991wu,Bassetto:1984ik,Ellis:1991qj} and in the in-medium antenna \cite{MehtarTani:2010ma,CasalderreySolana:2011rz,MehtarTani:2011gf,MehtarTani:2012cy,Fickinger:2013xwa}.

The gluon spectrum is composed of the direct emissions of the initial and final state and the interferences between both emitters.  Gluon emissions associated to the initial state contain  the $p_T$-broadening of the collinear gluon emitted before the hard scattering. The final state radiation corresponds to the genuine medium-induced gluon radiation off a highly energetic parton created inside the medium, so it is identified as the BDMPS-Z + vacuum spectrum.  In the case of the interferences,  the scale which measures the colour correlation between the initial and final state is the time-dependent  transverse size of the outgoing quark and the gluon radiated by the initial state prior to the hard scattering i.e. the quark-gluon system which traverses the QCD medium.  The evolution of the quark-gluon system when crossing the medium is accounted for by the propagator $\KO$. Depending on the asymptotic behaviour of the argument of the phase of $\KO$ we have two extreme cases: the coherent ($\tau\gg L^+$) and the incoherent ($\tau_f\ll L^+$) limits. 

Depending on the interplay of the kinematical variables we have a partial suppression of interferences in the coherent regime. Interferences vanish completely when $|\delta\k|\ll Q_s$ so there are gluon emissions outside of the angle associated to the hard scattering $\theta_{qq}$ and hence, antiangular ordering. The typical momentum scale transferred by the medium $Q_s^2=\hat q L^+$ sets up a upper bound for large angle radiation, so angular emissions lie in the range $\theta_{qq}\leq\theta\leq\theta_{max}$ where $\theta_{max}\sim Q_s/k^+$. The spectrum is completely suppressed above $\theta_{max}$ and the vacuum coherence pattern is recovered. When $|\delta\k|\gg Q_s$ gluons are radiated in a coherent manner as in the vacuum. 

In the incoherent regime we observed also a partial suppression of interferences. The spectrum in this case consists of the two emitters associated to either the initial or final state and an interference term which is developed at early times in the medium. Due to the multiple scatterings in the medium both emitters are able to radiate gluons at large angle emissions in the kinematic region $\theta_{qq}\leq\theta\leq \theta_{max}$ and the vacuum coherence pattern is reestablished  for $|\k|>Q_s$ due to the non-vanishing interference term. All these findings are in agreement with the corresponding results in the antenna \cite{MehtarTani:2010ma,CasalderreySolana:2011rz,MehtarTani:2011gf,MehtarTani:2012cy}.

We also discussed how the multiplicity of gluons is determined by the scale that dominates the process. We provide specific formulae for particle production in those situations. The observed logarithmic modification is a strong hint that the presence of a finite size QCD dense medium is indeed affecting the perturbative evolution of the PDFs and fragmentation functions. We leave the study of such modifications for a future work. 

On the phenomenological side, we expect that our study has consequences on particle production in nuclear collisions in the forward and backward regions i.e. for particle production associated to jets produced at angles different from $\pi/2$ with respect to the collision axis where the interferences between initial and final state radiation are absent. More specifically, we expect a difference in jet substructure for those jets going in the forward or backward direction from those going at $\pi/2$. This difference does not come from a different medium seen at different rapidities but from the appearance of medium-controlled interferences between initial and final state radiation. While the picture of the QCD medium produced at rapidities different from central is still unclear, the large rapidity plateau in inclusive particle production at the LHC gives the hope that the medium may not be dramatically different from the one at mid-rapidities, even at pseudorapidities such that the angle is sizeably different from $\pi/2$.

As an outlook, we plan to pursue the issue of which kind of factorisation may be valid in the regime where usual collinear factorisation plausibly fails, and the connection with other formalisms e.g. the hybrid formalism \cite{Dumitru:2005gt,Altinoluk:2011qy}. We also plan to examine the generalisation of our results when small-$x$ evolution of the target is allowed.

\section*{Acknowledgements}

We thank T. Altinoluk,  G. Beuf and K. Tywoniuk for their valuable comments and useful discussions. The work of NA, HM, MM and CAS is supported by European Research Council grant HotLHC ERC-2011-StG-279579; by Ministerio de Ciencia e Innovaci\'on of Spain under projects FPA2008-01177, FPA2009-06867-E and FPA2011-22776; by Xunta de Galicia (Conseller\'{\i}a de Educaci\'on and Conseller\'\i a de Innovaci\'on e Industria - Programa Incite); by the Spanish Consolider-Ingenio 2010 Programme CPAN and by FEDER. The work of YMT is supported by the European Research Council under the Advanced Investigator Grant ERC-AD-267258. NA, MM and YMT acknowledge financial support from ECT* during the workshop {\it h3QCD (high energy, high density and hot QCD)} where parts of this work were completed.

\appendix

\section{General form of the medium-induced gluon spectrum}
\label{sec:spec-noav}
The general functional form of the components of the gluon spectrum is given by
\begin{subequations}
\label{eq:gluoncomp-1}
\beq
\label{eq:befbef-1}
k^+ \frac{dN_{bef-bef}}{d^3k}&=& \frac{\alpha_s\,C_F}{\pi^2}\,\text{Re}\Biggl\{ \int d^2\r d^2\r' d^2\l d^2\l' \frac{d^2\pp}{(2\pi)^2}  \frac{d^2\pp'}{(2\pi)^2}e^{i k^+\u\cdot(\l-\l')} e^{-i\k\cdot(\r-\r')}e^{i(\pp\cdot\l-\pp'\cdot\l')}
\nonumber\\
&\times&  \frac{\pp\cdot\pp'}{\pp^2\,\pp'^2}\frac{\bigl\langle \mathcal{G}\bigl(L^+,\r,0,\l| k^+ \bigr) \mathcal{G}^\dagger\bigl(L^+,\r',0,\l'| k^+ \bigr)\bigr\rangle}{N_c^2-1}\Biggr\}\ ,\\
\label{eq:inin-1}
k^+ \frac{dN_{in-in}}{d^3k}&=& \frac{\alpha_s\,C_F}{2\pi^2}\frac{1}{(k^+)^2}\,\text{Re}\Biggl\{\int_0^{L^+} dy^+\int_0^{y^+} dy'^+ \int d^2\z d^2\z' e^{-i\k\cdot (\z-\z')}e^{i k^+\bar{u}^-(y^+-y'^+)}\nonumber \\
&\times& \bigl( i\partial_\y+k^+\bar{\u}\bigr)\cdot\bigl(- i\partial_{\y'}+k^+\bar{\u}\bigr) \\
&\times&\frac{\bigl\langle \mathcal{G}\bigl(L^+,\z,y^+,\y=\bar{\u}y^+| k^+ \bigr)\mathcal{U}(y^+,y'^+) \mathcal{G}^\dagger\bigl(L^+,\z',y'^+,\y'=\bar{\u}y'^+| k^+ \bigr)\bigr\rangle}{N_c^2-1}
\Biggr\}\ , \nonumber\\
\label{eq:inout-1}
k^+ \frac{dN_{in-out}}{d^3k}&=&  \frac{\alpha_s\,C_F}{\pi^2}\frac{1}{k^+}
\,\text{Re}\Biggl\{ 
i e^{i(k^--k\cdot\bar{\u})L^+}\int_0^{L^+} dy^+\int d^2\z e^{i (k^+\bar{u}^-y^+-\k\cdot\z)} \nonumber\\
&\times&\frac{\bbkappa\cdot(i\partial_{\y}+k^+\u)}{\bbkappa^2}
\frac{\bigl\langle \mathcal{G}\bigl(L^+,\z,y^+,\y=\bar{\u}y^+| k^+ \bigr) \mathcal{U}^\dagger\bigl(L^+,y^+\bigr)\bigr\rangle}{N_c^2-1}
\Biggr\}\ ,\\
\label{eq:outout-1}
k^+ \frac{dN_{out-out}}{d^3k}&=&\frac{\alpha_s\,C_F}{\pi^2}\,\frac{1}{\bbkappa^2}\ ,\\
\label{eq:befin-1}
k^+ \frac{dN_{bef-in}}{d^3k}&=& \frac{\alpha_s\,C_F}{\pi^2}\frac{1}{k^+}
\text{Re}\Biggl\{i \int_0^{L^+}dy^+\int \frac{d^2\k'}{(2\pi)^2}\, d^2\r\, d^2\z\, d^2\l\, e^{-ik^+\bar{u}^-y^+}e^{i \l\cdot(\k'+k^+\bar{\u})}e^{-i\k\cdot(\r-\z)}
\nonumber \\
&&\times\frac{\k'\cdot (-i\partial_{\y}+k^+\bar{\u})}{\k'^2}\nonumber \\
&&\times
\frac{\bigl\langle \mathcal{G}\bigl(L^+,\r,0,\l| k^+ \bigr) \mathcal{U}^\dagger\bigl(y^+,0\bigr)\mathcal{G}^\dagger\bigl(L^+,\z,y^+,\y=\bar{\u}y^+| k^+ \bigr) \bigr\rangle}{N_c^2-1}
\Biggr\}\ ,\\
\label{eq:befout-1}
k^+ \frac{dN_{bef-out}}{d^3k}&=&-2\frac{\alpha_s\,C_F}{\pi^2}\, 
\text{Re}\Biggl\{ e^{iL^+(k^--\k\cdot\bar{\u})}\int \frac{d^2\k'}{(2\pi)^2}\, d^2\r\, d^2\l\, e^{-i\k\cdot\r}e^{i \l\cdot(\k'+k^+\u)}\nonumber\\
&\times&\frac{\k'\cdot\bbkappa}{\bbkappa^2\k'^2}\frac{\bigl\langle \mathcal{G}\bigl(L^+,\r,0,\l| k^+ \bigr) \mathcal{U}^\dagger\bigl(L^+,0\bigr)\bigr\rangle}{N_c^2-1}
\Biggr\}\ .
\eeq
\end{subequations}
Using the harmonic oscillator approximation, one simply reduces to Eqs. (\ref{eq:gluoncomp-2}).

\section{Vacuum emission pattern between the initial and final state radiation}
\label{sec:vac-spec}
In momentum space, the total current $\J^{\mu}_{(0)}=\J^{\mu}_{bef,(0)}+\J^{\mu}_{aft,(0)} $, where $\J^{\mu}_{bef,(0)}$ and $\J^{\mu}_{aft,(0)}$ are given by Eqs. (\ref{eq:inc-outcurr}), can be written as
\beq
\label{eq:totvaccurr}
J_{(0),a}^{\mu}=ig\left( -\frac{p^\mu}{p\cdot k +i\epsilon}Q_{bef}^a\ +\ \frac{\bar p^\mu}{\bar p\cdot k+i\epsilon}Q_{aft}^a\right) \,.
\eeq
By linearizing the CYM equations, the solution of the classical gauge field at leading order in momentum space reads
\beq
\label{eq:solvac}
-k^2 A_{(0)}^{i,a}=\, 2ig \left( \frac{\kappa^i}{\bkappa^2}Q_{in}^a - \frac{\bar\kappa^i}{\bbkappa^2}Q_{out}^a\right) \,.
\eeq
By replacing this solution of the gauge field into the reduction formula (\ref{eq:redform-a}) and summing over the physical polarisations, the vacuum spectrum of the inclusive gluon spectrum for the singlet case results
\beq
\label{eq:vacspec}
\omega\frac{dN^\text{vac}}{d^3\vec{k}}=\frac{\alpha_s  C_F}{\pi^2}\,
\Biggl( \frac{1}{\bkappa^2}+\frac{1}{\bbkappa^2}-2\frac{\bkappa\cdot\bbkappa}{\bkappa^2\bbkappa^2}
\Biggr)\,,
\eeq
which is precisely the vacuum gluon spectrum \cite{Dokshitzer:1991wu,Bassetto:1984ik,Ellis:1991qj}.

\section{Derivation of the ``\emph{in-in}" contribution, Eq. (\ref{eq:inin-lead})}
\label{sec:ininapp}

In this section we show how to get the correct leading order result of the ``\emph{in-in}" component, Eq. (\ref{eq:inin-lead}). We follow closely the procedure worked out in Ref. \cite{MehtarTani:2012cy}. The phase of tangent function in Eq. (\ref{eq:inmed}) goes roughly like $|\Omega| y^+\sim y^+/\tau_f$. Therefore we can divide in Eq. (\ref{eq:inmed}) the integration interval over $y^+$ in two parts, the one related with early-time emissions where $0\leq y^+\ll \tau_f$ and the other one for late-time emissions when $\tau_f\ll y^+\lesssim L^+$. To make a clear distinction between these two regimes, let us introduce a dimensionless real number $c\gg 1$ such that $c\tau_f\ll L^+$. We will see that the leading term of the ``\emph{in-in}" component does not depend on this parameter \cite{MehtarTani:2012cy}. 

If we consider first the case of late-time emissions $c \tau_f\ll y^+$, Eq.(\ref{eq:inmed}) can be approximated by
\beq
\label{eq:inin-late}
k^+ \frac{dN_{in-in}}{d^3k}\Biggl.\Biggr|_{y^+\gg c\tau_f}&=&\frac{\alpha_s\,C_F}{k^+ \pi^2}\int_{c\tau_f}^{L^+}dy^+\int \frac{d^2\k}{(2\pi)^2}\P(\k-\bbkappa,L^+-y^+)\nonumber \\
&\times& e^{-\frac{\k^2}{k_f^2}}\sin\Biggl(\frac{\k^2}{k_f^2}\Biggr)\,.
\eeq
For early-time emissions we can approximate $\P(\k-\bbkappa,L^+-y^+)\approx \P(\k-\bbkappa,L^+)$ and we rewrite Eq. (\ref{eq:inmed}) as
\beq
\label{eq:inin-early}
k^+ \frac{dN_{in-in}}{d^3k}\Biggl.\Biggr|_{y^+\ll c\tau_f}&=&\frac{\alpha_s\,C_F}{2(k^+\pi)^2}\text{Re}\Biggl\{
\int \frac{d^2\k}{(2\pi)^2}\P(\k-\bbkappa,L^+)\nonumber \\
&\times&\int_{0}^{\beta}dx\,\frac{2ik^+}{\Omega}\exp\Biggl[(1-i)\frac{\k^2}{k_f^2}\tan[x]
\Biggr]
\Biggr\}\,,
\eeq
where $\beta=(1-i)c/2$. To perform the integral over time, we make use of the exact integral
\beq
\label{eq:exp-integral-def}
\int dx \,e^{a \tan x} = \frac{i}{2}\left\{ e^{ia}E_1 \left[a (i-\tan x) \right] - e^{-ia} E_1\left[-a (i + \tan x) \right] \right\} \,+\text{const.},
\eeq
where $E_1(x)= - Ei(-x)$ is the exponential integral. If we keep just the leading term in $c$ we can approximate $\tan\beta\approx -i + 2ie^{-2i \beta}$, so finally 
\beq
\label{eq:exp-integral}
\int_0^{\beta} dx \exp \left[ (1-i) \frac{\k'^2}{2 k_f^2} \tan(x)\right]  &\simeq &  -\frac{i}{2} \Big\{ e^{ia}\big[ E_1(ia)-E_1(2ia) \big] \nn
&-&e^{-ia} \big[ E_1(-ia) + \gamma_E + \ln(-2ia) - 2i \beta \big] \Big\},
\eeq
where now $a \equiv (1-i) \k'^2/(2 k_f^2)$ and we have expanded the last term using $E_1(x)\simeq -\gamma_E-\ln(x)$ for $x \ll1$. The last term in Eq.~(\ref{eq:exp-integral}) gives rise to a term of the form of the Eq.~(\ref{eq:inin-late}), however with $y^+$ running from 0 to $c\tau_f$. Adding it to Eq.~(\ref{eq:inin-late}) removes the spurious dependence on $c$, as we pointed out above. The remaining terms are not proportional to $L^+$ so they can be neglected at this level of the approximation. This shows that the leading order term of the ``\emph{in-in}"  component is effectively captured by Eq. (\ref{eq:inin-lead}). Notice that if one considers $\k^2 > k_f^2$ in Eq. (\ref{eq:exp-integral}) it is possible to get the leading terms of the exponential integral $E_1(x)\approx e^{-x}/x$ so we obtain
\beq
\label{eq:exp-integral-2}
\frac{4i\omega}{\Omega}\int_{0}^{\beta}dx \exp\Biggl\{(1-i)\frac{\k^2}{k_f^2}\tan x
\Biggr\}\Biggl.\Biggr|_{\k^2>k_f^2}\simeq \frac{8\omega^2}{\k^2}\,;
\eeq
additional terms are suppressed as $\exp[-\k^2/k_f^2]$ and thus can be neglected. This early-time collinear divergence arises from the bremsstrahlung emission at early times which is followed by rescattering of the emitted gluon in the medium.  This collinear divergence is balanced out by the ``\emph{in-out}" and ``\emph{out-out}" contributions to get effectively the genuine vacuum contribution.  A complete and detailed analysis of this subtle aspect is found in Sect. 5 of Ref. \cite{MehtarTani:2012cy}. 

\bibliography{biblio}{}
\bibliographystyle{jhep}

\end{document}